\documentclass[journal]{IEEEtran}

\usepackage{amsmath}
\usepackage{amssymb}
\usepackage{mathrsfs}
\usepackage{cancel}
\usepackage{footmisc}
\usepackage{color,soul}
\usepackage{xcolor}
\usepackage{colortbl}
\usepackage{hyperref}
\usepackage{cleveref}
\usepackage{cite}
\usepackage{siunitx}
\usepackage[version=4]{mhchem}
\usepackage{algorithm}
\usepackage{algpseudocode}
\usepackage{booktabs}
\usepackage{tikz}
\usepackage{pgfplots}
\usetikzlibrary{shapes.geometric, arrows.meta, positioning, fit, calc, tikzmark}
\pgfplotsset{compat=newest}
\usepackage{circuitikz}
\usepackage{bm}
\usepackage{subcaption}
\usepackage{stfloats}
\usepackage{graphicx, overpic} 

\newcommand{\sectionrefname}{\textit{Section }}
\usepackage{caption}
\DeclareSIUnit{\MVA}{MVA}

\newcommand{\redY}{ \mathcolor{orange}{\mathbf{y}}}
\newcommand{\redy}{ \mathcolor{orange}{y}}
\newcommand{\blueU}{\mathcolor{blue}{\mathbf{u}}}
\newcommand{\blueu}{\mathcolor{blue}{u}}

\newcommand{\tikzsquare}{%
  \begin{tikzpicture}[baseline=0.4em]
    \filldraw[fill=white, draw=black] (0,0) rectangle (1.4em, 1.4em);
  \end{tikzpicture}%
}
\newcommand{\tikztriangleLow}{%
  \begin{tikzpicture}[baseline=0.4em]
    \filldraw[fill=white, draw=black] (0,0) -- (1.4em, 0) -- (0, 1.4em) -- cycle;
  \end{tikzpicture}%
}

\newcommand*{\horzbar}{\rule[.5ex]{2.5ex}{0.5pt}}

\def\BibTeX{{\rm B\kern-.05em{\sc i\kern-.025em b}\kern-.08em
    T\kern-.1667em\lower.7ex\hbox{E}\kern-.125emX}}
\begin{document}

\title{Data-Driven Predictive Control for \\Wide-Area Power Oscillation Damping}

\author{Giacomo~Mastroddi,
        Jan~Poland,
        Mats~Larsson,
        and~Keith~Moffat%
\thanks{G. Mastroddi is with ETH Zurich, Zurich, Switzerland (e-mail: giacomma@ethz.ch).}%
\thanks{J. Poland and M. Larsson are with Hitachi Energy Research, Baden-Dättwil, Switzerland (e-mail: jan.poland@hitachienergy.com; mats.larsson@hitachienergy.com).}%
\thanks{K. Moffat is with The University of Melbourne, Melbourne, Australia (e-mail: keith.moffat@unimelb.edu.au).}%
\thanks{This work has been submitted to the IEEE for possible publication. Copyright may be transferred without notice, after which this version may no longer be accessible.}
}


\maketitle

\begin{abstract}
We study damping of inter-area oscillations in transmission grids using voltage-source-converter-based high-voltage direct-current (VSC-HVDC) links. Conventional power oscillation damping controllers rely on system models that are difficult to obtain in practice. 
Data-driven Predictive Control (DPC) addresses this limitation by replacing explicit models with data. 
We apply AutoRegressive with eXogenous inputs (ARX)–based predictive control and its Transient Predictive Control (TPC) variant, and compare them with Data-enabled Predictive Control (DeePC) and two standard model-based controllers. 
The methods are evaluated in simulation on a system exhibiting both inter-area and local oscillation modes. 
ARX-based predictive control and DeePC both achieve effective damping, while the ARX-based methods require less online computation. 
Using warm-started, pre-factorized operator-splitting solvers, ARX/TPC control actions are computed in less than 1~ms. 
These results demonstrate that DPC is a viable approach for power-system oscillation damping for the given test case.


\end{abstract}

\begin{IEEEkeywords}
data-driven control, predictive control, online optimization, ARX, HVDC, wide-area control, power oscillation damping, inter-area oscillations
\end{IEEEkeywords}

\section{Introduction}
\label{sec:introduction}
Large interconnected power grids exhibit low-frequency inter-area oscillations when electromechanical modes coupling groups of synchronous generators (SGs) across areas become weakly damped, often under high inter-area transfers over weak tie-lines (e.g., long, high-impedance corridors) \cite{kundur1994, winkelman1981analysis}.
The ongoing energy transition, involving increased renewable integration, may increase the occurrence of such oscillations as conventional SGs equipped with power system stabilizers (PSSs) are phased out \cite{2016oscillations, gu2022power, dorfler2023control, zhao2023comparison, krajacic2025revisiting}.
As the damping contribution from PSSs decreases, converter-based solutions, such as  High-Voltage Direct Current (HVDC) control, can fill the grid-stabilization gap.  



Inter-area oscillations have traditionally been addressed using generator-side PSSs, however, effective coordination of multiple PSSs is challenging due to parameter uncertainty and operating-point dependence \cite{kundur1994, rogers2000}.
Recent case studies support this point, showing that poorly tuned or uncoordinated PSSs can lead to small-signal instability \cite{krajacic2025revisiting, alghamdi2025pso}.
Transmission-side solutions based on Flexible Alternating Current Transmission Systems (FACTS) devices \cite{hingorani2000}, including series controllers \cite{Hingorani1993} and shunt devices such as Static Var Compensators (SVCs) and Static Synchronous Compensators (STATCOMs) \cite{rogers2000, larsen1995}, 
have also been deployed, though their impact on inter-area modes is state- and location-dependent.
More recently, wide-area measurement-based controllers have been proposed to enhance damping robustness, at the cost of increased communication and implementation complexity \cite{uhlen2012wide, yao2014wide, schoenwald2017design, zhao2023comparison}.

HVDC damping is compelling because an HVDC link connects to the AC grid at two geographically separated terminals, enabling wide-area actuation.
The advent of voltage-source-converter (VSC) technology enabled HVDC links to modulate power faster than line-commutated converters, making HVDC-based inter-area oscillation damping possible \cite{imhof2015voltage, elizondo2018interarea, cowan2021_hvdc_converter_choice_gb}. 
This capability has been studied in the academic literature, developing model-based methods such as Modal Residues (MR) and frequency-domain Loop Shaping (LS) \cite{rogers2000, gibbard2015, roberson2017_loop_shaping_hvdc}. 
Despite this maturity, such controllers have seen limited adoption in industry and the potential of HVDC power oscillation damping remains underutilized, largely because their performance depends on accurate linearized system and converter models \cite{elizondo2018interarea, xing2021advanced}.

Adaptive and data-driven control provides an alternative to model-based designs.
Adaptive oscillation damping has a long history in power systems, focusing on PSSs \cite{hassan1993_fuzzy_pss, chen1993_apss_poleshift} in the 1990s and FACTS \cite{chaudhuri2004_mmac, sadikovic2006_selftuning_facts} in the 2000s. 
Recently, wide-area oscillation damping has focused on VSC-HVDC links as actuators \cite{elizondo2018interarea, dong2023_adaptive_pod_vschvdc_gb}.
Some methods learn (local) linear models of the system, \cite{bai2016design, zhu2018_ringdown_tf_wadc, shi2020, han2022model}, while others apply Reinforcement Learning (RL) to learn damping controllers directly \cite{shen2018, hashmy2020_wams_rl_lfo_damping, huang2022, zhang2024}.

While RL-based damping exists in the literature, Model Predictive Control (MPC) frameworks \cite{azad2013damping, fuchs2013stabilization, fan2018nonlinear} more naturally support hard constraints.
MPC deployments, on the other hand, are challenging given their sensitivity to model fidelity and the need to solve optimization problems online.


Data-driven Predictive Control (DPC) methods \cite{favoreel1999spc, berberich2020data, dorfler2022bridging, breschi2023data} such as Data-enabled Predictive Control (DeePC) \cite{coulson2019dataenabledpredictivecontrolshallows} and AutoRegressive with eXogenous inputs (ARX)-based methods \cite{LjungBook, LjungMcK1996} such as Transient Predictive Control (TPC) \cite{2024tpc} do not require a system model and promise plug-and-play installation. 
Moreover, DPC methods naturally apply to Multiple-Input, Multiple-Output (MIMO) settings with input and output constraints. 

A limitation of DeePC and ARX-based methods is that they are derived assuming that the system is Linear, Time-Invariant (LTI). 
Power systems are nonlinear, and thus it is not obvious that LTI-based DPC methods can be effective for grid stabilization. 
This paper investigates if LTI-based DPC methods can be effective for damping power system oscillations by adjusting the power and voltage setpoints of the VSC-HVDC converters that interface with the AC grid. 

DeePC \cite{coulson2019dataenabledpredictivecontrolshallows} is based on Willems’ Fundamental Lemma \cite{willems2005} and Subspace and Behavioral Systems Theory \cite{markovsky2023data}. It has been applied to power systems applications \cite{markovsky2023data} and shown to outperform MPC based on Auto-Regressive-Moving-Average (ARMA) models identified via prediction error methods \cite{huang2019dataenabledpredictivecontrolgridconnected}. 
Subsequent work, investigating a similar HVDC control problem to this work, extended DeePC to decentralized wide-area control \cite{huang2021decentralizeddataenabledpredictivecontrol}. 
A DeePC variant, velocity-DeePC, was developed in \cite{2023felix}, which investigates data-driven inverter control for power-system applications where training data are gathered online under closed-loop feedback.

TPC, recently proposed in \cite{2024tpc}, offers an alternative data-driven approach to subspace-based methods \cite{favoreel1999spc, coulson2019dataenabledpredictivecontrolshallows}, which compresses the training data offline using the ``Transient Predictor.'' TPC's causal and unbiased predictions in closed-loop settings make it attractive for power system applications \cite{moffat2025biassubspacebaseddatadrivenpredictive}. Also, like DeePC and other ARX-based methods, it can readily handle MIMO systems. Implementation for grid-connected inverters has been presented in \cite{graf2025gridconnecteddatadriveninvertercontrol}. 

TPC is an ARX-based controller, and is an alternative to ``Fixed-Length Predictor'' described in \cite{2024tpc}. 
In this paper, we use the terminology \emph{Single-ARX Control} with the understanding that it is synonymous to \emph{Fixed-Length Predictor}.
We benchmark Single-ARX Control alongside TPC.

This paper contributes to the practical deployment of DPC methods for inter-area oscillation damping. Specifically, we demonstrate that LTI-based DPC methods can be used to control a VSC-HVDC line to mitigate synchronous oscillations without a system model. 
We conduct our experiments on an industry-standard, high-fidelity, nonlinear four-area Kundur benchmark system \cite{huang2021decentralized, 2023felix}, including additive process noise and dynamic models of the HVDC link.
We find that the DPC methods perform comparably well with industry-standard MR and LS methods \cite{gibbard2015}, which rely on accurate linearized power-system and HVDC models.

Two primary concerns with predictive control for power systems are 1) model fidelity and 2) optimization cost.
To assess the viability of LTI-based DPC control for power system stabilization, we investigate the linearity of the power system from the perspective of the VSC-HVDC ports.
To assess the computational burden of DPC, we tested DeePC and TPC with four open-source convex optimization solvers---two Interior-Point methods, Clarabel \cite{clarabel} and PIQP \cite{piqp}---and two Operator-Splitting methods, OSQP \cite{osqp} and SCS \cite{scs2016,scs2021}. The results, including a comparison between DPC methods, are reported. 

Summarizing, the contributions of this paper are:
\begin{enumerate}
    \item a linearity analysis of the test case that supports the use of linear DPC tools, 
    \item a computational comparison between three DPC methods and between four open source convex optimization solvers, and
    \item a proof-of-concept that the three DPC methods perform comparably to the model-based tools on the case study.
\end{enumerate}

The remainder of this paper is organized as follows. \sectionrefname\ref{sec:system} introduces the benchmark grid and presents the linearity analysis. \sectionrefname\ref{sec:lead_lag} presents the model-based industry-standard controllers. \sectionrefname\ref{sec:DPC} introduces the DeePC, TPC, and Single-ARX Control architectures. \sectionrefname\ref{sec:results} gives the computational and control performance results. Conclusions are summarized in  \sectionrefname\ref{sec:conclusion}.

\begin{figure*}[t]
    \centering
    \input{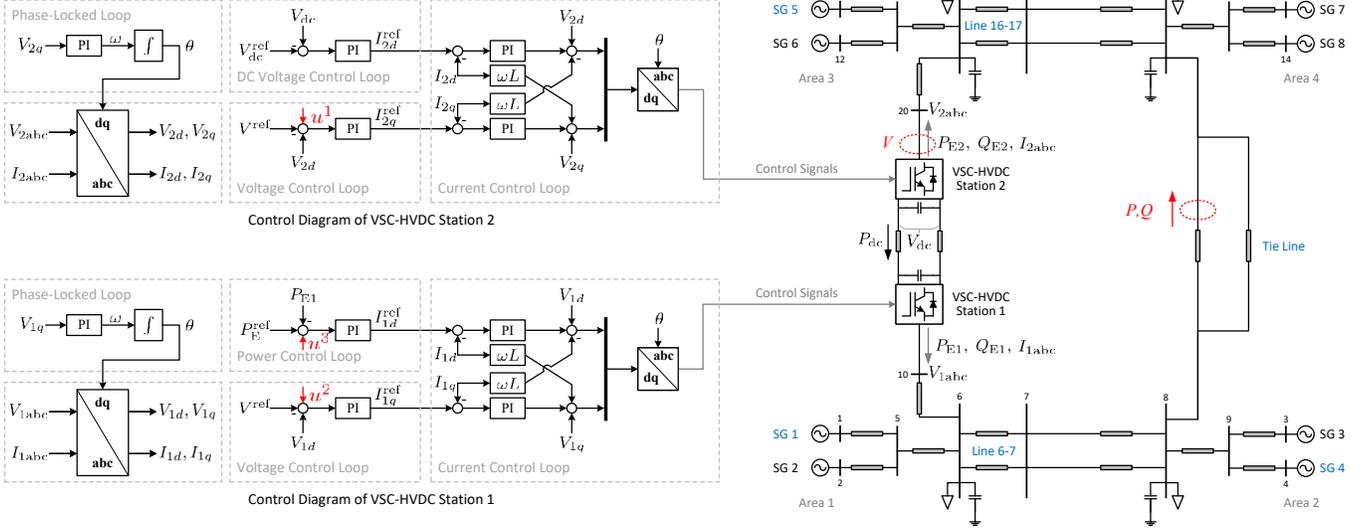}
    \caption{One-line diagram of the four-area power system with HVDC link. The control inputs $(u^1, u^2, u^3)$ and outputs $(V, P, Q)$ are in red.}
    \label{fig:power_system}
\end{figure*}

\section*{Notation}
In this paper, the number of data samples used to construct the Hankel matrix for 
the data-driven models is $n_\mathrm{samples}$ and the number of rows and columns of the Hankel matrix 
are $n_\mathrm{row}$ and $n_\mathrm{col}$, respectively. For any signal $\textbf{w}(t) \in \mathbb{R}^{n_w}$, we define the associated $\frac{1}{\sqrt{n_\mathrm{col}}}$-scaled Hankel matrix $W_{[t_0,t_1]} \in \mathbb{R}^{n_\mathrm{row} \times n_\mathrm{col}}$ with $n_\mathrm{row} = n_w (t_1 - t_0 + 1)$ as:
\begin{equation*}
\resizebox{\linewidth}{!}{$
    \mathcal{H}(\textbf{w})_{[t_0, t_1], n_\mathrm{col}} = \frac{1}{\sqrt{n_\mathrm{col}}} 
    \begin{bmatrix} 
        \textbf{w}(t_0) & \textbf{w}(t_0 + 1) & \dots & \textbf{w}(t_0 + n_\mathrm{col} - 1) \\ 
        \textbf{w}(t_0 + 1) & \textbf{w}(t_0 + 2) & \dots & \textbf{w}(t_0 + n_\mathrm{col}) \\ 
        \vdots & \vdots & \ddots & \vdots \\ 
        \textbf{w}(t_1) & \textbf{w}(t_1 + 1) & \dots & \textbf{w}(t_1 + n_\mathrm{col} - 1) 
    \end{bmatrix},
$}
\end{equation*}
where the $\frac{1}{\sqrt{n_\mathrm{col}}}$ scaling normalizes the variance. Vectors are lower-case and bold, while matrices are upper-case. Numeric superscripts $i$ indicate the entry of a vector, such as the $i^\mathrm{th}$ input or output, while the subscripts are used for time stamps or names.

\section{Test Case: Four-area Power System}
\label{sec:system}
The benchmark power grid used throughout this paper is 
a four-area model which consists of two copies of the well-known benchmark from \cite{kundur1994} that have been connected with a long Alternating Current (AC) transmission line as well as an HVDC link. This system exhibits multiple oscillatory modes over a broad frequency range, making it a demanding test case, close to real world application. \autoref{fig:power_system} shows the system layout, where the bottom and top two areas are interconnected via a \SI{1000}{\MVA} VSC-HVDC link in parallel with an AC line. The simulation model is implemented with full Electromagnetic Transient dynamics and detailed HVDC dynamics, including the Modular Multilevel Converter dynamics.

We approximate successive stages of the energy transition, from a conventional fossil-based system to one dominated by renewable sources, by progressively deactivating PSSs across an increasing number of SGs, as summarized in \autoref{tab:cases}. 

\begin{table}[h!]
\caption{Simulation cases representing successive stages of the energy transition, defined by the percentage of deactivated PSSs.}
\label{tab:cases}
\centering
\begin{tabular}{S c l}
    \toprule
    \textbf{PSSs Deactivated [\%]} & \textbf{\autoref{fig:openloop_modes}} & \textbf{System Description} \\ 
    \midrule
    0  & $\times$             & Traditional operation \\
    25 & $\color{red}{\circ}$ & Early transition stage \\
    50 & $\color{blue}{*}$    & Intermediate stage \\
    75 & $\color{magenta}{+}$ & High renewable share \\
    \bottomrule
\end{tabular}
\end{table}

Each SG is initially equipped with a PSS tuned according to the parameters in~\cite{kundur1994}, ensuring adequate damping of all modes when fully active (\SI{0}{\percent} PSSs deactivated).

\begin{figure}[t!]
    \centering
    \includegraphics[width=0.72\linewidth]{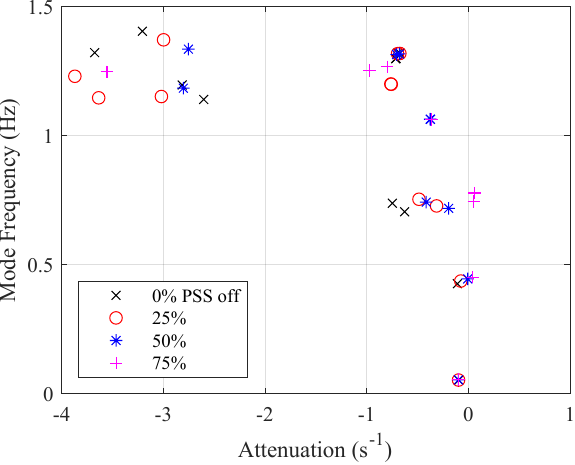}
    \caption{Open-loop pole locations with varying amounts of active PSSs.}
    \label{fig:openloop_modes}
\end{figure}

\autoref{fig:openloop_modes} displays the open-loop pole locations of the low-frequency modes below~\SI{2}{\hertz} for different levels of PSS deactivation. As the share of inactive PSSs increases, inter-area modes become progressively less damped until the system turns open-loop unstable for \SI{75}{\percent} PSSs deactivated, due to negatively damped inter-area modes.

Inspection of the eigenvalue spectrum reveals one dominant inter-area mode at~\SI{0.44}{\hertz}, two additional inter-area modes around~\SI{0.77}{\hertz}, and local modes near~\SI{1.3}{\hertz}. Further analysis using mode shapes indicates that the~\SI{0.44}{\hertz}~mode relates to SGs in Areas~1-2 oscillating as a coherent group against the SGs in Areas~3-4, while the two~\SI{0.77}{\hertz}~modes correspond to oscillations of Area~1 against Area~2 and of Area~3 against Area~4.

\subsection{Proposed Control Inputs}
We propose an outer-level control that is compatible with the existing converter control structure and does not require altering or retuning the internal controllers. The proposed damping control operates by making adjustments to the selected reference signals of the VSC-HVDC control system. This control structure can provide wide-area damping without interfering with normal operation.

HVDC lines permit three degrees of freedom, and thus three adjustment signals/inputs are used ($m = 3$): the voltage reference at Station 2, and the voltage and active power references at Station 1. The input vector is defined as
\begin{equation*}
    \textbf{u} =
    \begin{bmatrix}
        u^1 \\ u^2 \\ u^3
    \end{bmatrix} = 
    \begin{bmatrix}
        \Delta V_\mathrm{ref}^2 \\ \Delta V_\mathrm{ref}^1 \\ \Delta P_\mathrm{ref}^1
    \end{bmatrix}.
\end{equation*}
To normalize the control task, each quantity is expressed in per unit (pu).
Each control signal is bounded by the HVDC converter ratings. We define the feasible input set as
\begin{equation*}
    \mathcal{U} = \{ \textbf{u} \in \mathbb{R}^m \mid -0.1 \leq u^i \leq 0.1, \forall i \in m\},
\end{equation*}
ensuring that the modulation of the references remains within $\pm \SI{0.1}{pu}$ of their nominal values.

\subsection{System Outputs}
\label{sec:outputs}
Following the configuration proposed in~\cite{2023felix}, three filtered signals in pu are used as outputs (\(p = 3\)): the filtered voltage at Station~2, and the filtered active and reactive power flows in the tie line. These signals form the output vector
\begin{equation*}
    \textbf{y} =
    \begin{bmatrix}
        y^1 \\ y^2 \\ y^3
    \end{bmatrix} = 
    \begin{bmatrix}
        V \\ P \\ Q
    \end{bmatrix}.
\end{equation*}
These outputs are influenced by exogenous events, like load variations, line trips or faults, that lie outside our control. Thus, we do not assert output constraints, i.e. \(\mathcal{Y} = \{\mathbf{y} \in \mathbb{R}^p \}\), as doing so could render the optimization problem infeasible.

\begin{figure}[htb]
    \centering
    \begin{tikzpicture}[
    node distance=1cm and 2cm,
    block/.style={rectangle, draw, thick, minimum height=0.7cm, minimum width=1.8cm, fill=white, align=center},
    arrow/.style={-Stealth, thick},
]

\node[block] (agent) at (0,0) {\small Controller};
\node[block] (env) at (3,0) {\small System};
\draw (2,-1) to[bandpass, n=bpf] ++(2,0);

\draw[arrow] (agent.e) -- (env.west) node[midway, above] {$\textbf{u}$};

\draw[arrow] (env.east) -- ++(1.5,0) node[pos=0.5, above] {$\tilde{\textbf{y}}$} coordinate[pos=0.5] (y_tilde);
\draw[arrow] (y_tilde) |- (bpf.e); 

\draw[arrow] (bpf.w) -| (agent.south) node[pos=0.25, above] {$\textbf{y}$};

\end{tikzpicture}





    \caption{Closed-loop with band-pass filtering and mean removal.}
    \label{fig:closed_loop}
\end{figure}
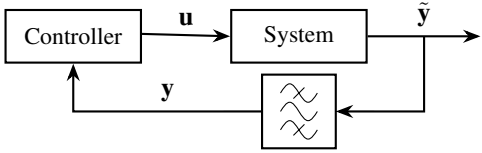

To ensure that the controller reacts only to the frequencies relevant for inter-area oscillations, the measured signals \(\tilde{\mathbf{y}}\) from the power system are processed through a band-pass filter with cutoff frequencies of \SI{0.05}{\hertz} and \SI{5}{\hertz}. In addition, the mean value of each signal is removed to focus exclusively on deviations from the steady-state operating point. This filtering isolates the modes of interest while suppressing both high-frequency dynamics and slow drifts.
The signal path from system measurements through filtering to control input generation is illustrated in~\autoref{fig:closed_loop}.

\subsection{Load Variations}
\label{sec:process_noise}
Process noise significantly impacts the robustness and effectiveness of controllers and must be considered. In the employed simulation model, process noise is modeled as random variations in each load in the system. A bank of low-pass filtered Pseudo-Random Binary Signals (PRBSs) with frequencies in the range of \SIrange{0.01}{10}{\hertz} are used to create a  mix of fast and slow load variations. This mix results in realistic voltage and frequency variations under stationary conditions.
Typical standard deviation of the load variations is set to \SI{0.01}{pu}, which results in standard deviation of frequency of around \SI{35}{\milli\hertz} when PSSs are active on all SGs.


\subsection{Data Sampling} 
The discrete-time simulation enforces strict causality, meaning that a control input applied at time $t_0$ affects the system outputs only from time $t_1$ onward.
Data is collected as paired Input-Output (I/O) vectors, and the system is sampled with a fixed interval of $T_\mathrm{s} = \SI{0.1}{\second}$.


\subsection{Linearity Analysis}
\label{sec:linearity_test}
As the DPC methods we investigate rely on linearity, we conduct a study to empirically evaluate how ``far'' the high-fidelity test case in \autoref{fig:power_system} is from linear by testing the additivity and homogeneity of the output of the nonlinear system.  

\begin{figure}[h!]
    \centering
    \includegraphics[width=0.6\linewidth]{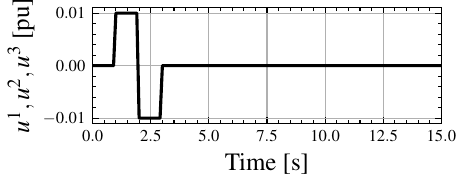}
    \caption{Single-Excitation Doublet used to excite the system.}
    \label{fig:linearity_base_input}
\end{figure}

To do this, we define a ``Single-Excitation Doublet'' as the step-increase step-decrease of only \textit{one} input while the other two inputs are held at zero. 
The Single-Excitation Doublets have amplitude \(\pm\SI{0.01}{pu}\) and duration \(\SI{1}{\second}\), as shown in \autoref{fig:linearity_base_input}.

We simulate the system with a Single-Excitation Doublet for each of the three inputs, producing the corresponding output vectors \(\bar{\mathbf{y}}_1\) for the $u^1$ Single-Excitation Doublet, \(\bar{\mathbf{y}}_2\) for the $u^2$ Single-Excitation Doublet, and \(\bar{\mathbf{y}}_3\) for the $u^3$ Single-Excitation Doublet. \(\bar{\mathbf{y}}_1\), \(\bar{\mathbf{y}}_2\), and \(\bar{\mathbf{y}}_3 \in \mathbb{R}^3\).
If the system is linear, then the output of a linear combination of the Single-Excitation Doublets will be equal to the corresponding linear combination of \(\bar{\mathbf{y}}_1\), \(\bar{\mathbf{y}}_2\), and \(\bar{\mathbf{y}}_3\).

\begin{figure}[t!]
    \centering
    \includegraphics[width=0.72\linewidth]{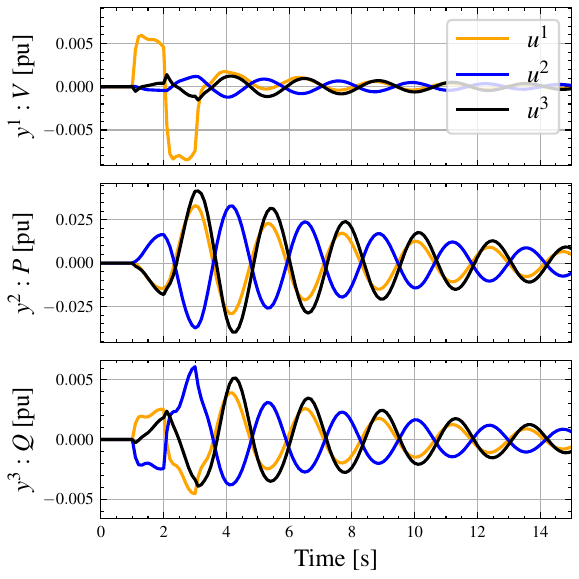}
    \caption{Output responses corresponding to the input signals in \autoref{fig:linearity_base_input}. The orange lines give the response when the $u^1$ Single-Excitation Doublet is applied ($u^2$ and $u^3$ are held constant at zero). The blue lines describe the $u^2$ Single-Excitation Doublet response and the black lines describe the $u^3$ Single-Excitation Doublet response.}
    \label{fig:linearity_base_output}
\end{figure}

\autoref{fig:linearity_base_output} describes the output of the Single-Excitation Doublet tests. Due to space considerations, \autoref{fig:linearity_base_output} combines the outputs of the Single-Excitation Doublet tests onto a single plot. In practice, all the orange lines were produced by the $u^1$ Single-Excitation Doublet test, all the blue lines by the $u^2$ test, and all the green lines by the $u^3$ test.

To conduct the linearity test we apply a linear combination of the Single-Excitation Doublets to the system. For example, we simulate the system with the input \(5(u^1 + u^2)\), where $u^1$ and $u^2$ are the Single-Excitation Doublet signals for inputs one and two, respectively. The resulting measured output is denoted by \(\mathbf{y}\). The linearly approximated output \(\hat{\mathbf{y}}\) is then constructed by scaling and summing the individual basis trajectories, i.e.
\(\hat{\mathbf{y}} = 5(\bar{\mathbf{y}}_1 + \bar{\mathbf{y}}_2)\).

System linearity is then quantified using the Root Mean Squared Error (RMSE) between the nonlinear output \(\mathbf{y}\) and its linear approximation \(\hat{\mathbf{y}}\) over the test's full time window:
\begin{equation*}
    \mathrm{RMSE} =
    \sqrt{
        \frac{1}{\tau_\mathrm{sim}}
        \lVert\hat{\mathbf{y}} - \mathbf{y}\rVert_2^2
    }.
\end{equation*}
A small RMSE indicates that the nonlinear simulation closely follows the expected linear behavior, supporting the use of linear data-driven models in subsequent analysis.


\begin{figure*}[t]
    \centering
    \includegraphics[width=0.75\linewidth]{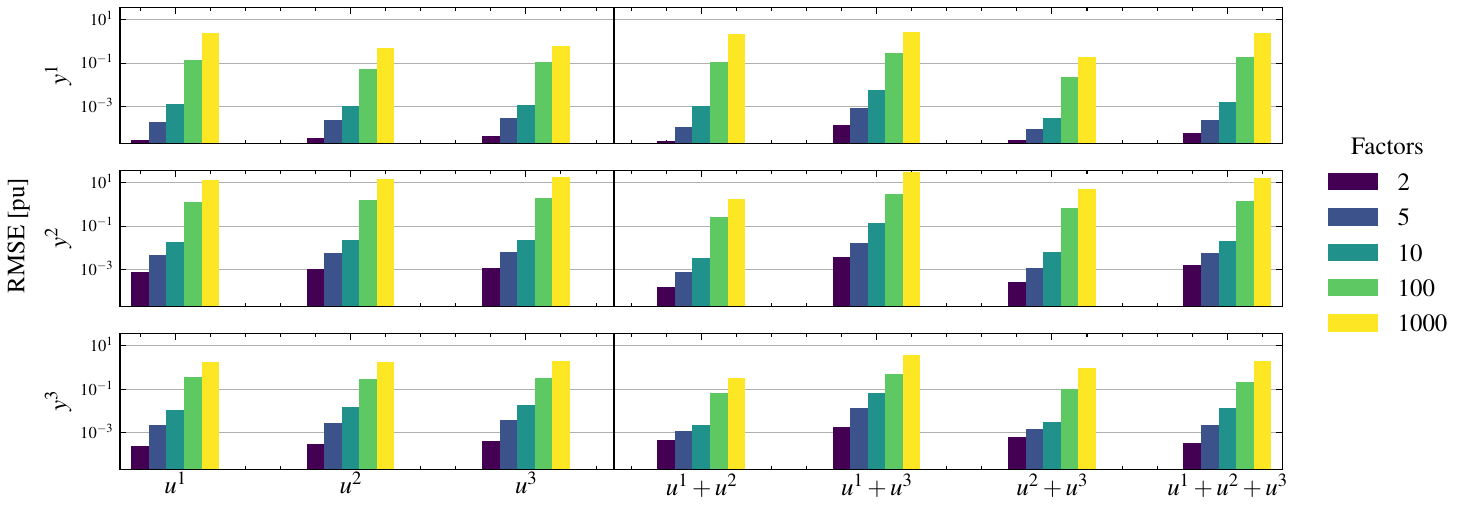}
    \caption{Linearity test results with \SI{50}{\percent} PSS deactivated with different scaling factors applied to the Single-Excitation Doublets for the three control inputs.}
    \label{fig:results_linearity}
\end{figure*}

\autoref{fig:results_linearity} shows the results of the linearity analysis. The horizontal axis is divided into two parts. On the left-hand side, the results correspond to tests of homogeneity only, where individual Single-Excitation Doublets are scaled by a constant factor. On the right-hand side, both homogeneity and additivity are tested by summing multiple Single-Excitation Doublets and applying the same scaling factor. \\
The Single-Excitation Doublets have a maximum amplitude of \(\pm\SI{0.01}{pu}\), while the control inputs are bounded at \(\pm\SI{0.1}{pu}\). Consequently, scaling factors up to 10 correspond to physically relevant operating conditions, whereas larger factors are included only to illustrate the limits of linear behavior. \\
The vertical axis shows the RMSE expressed in absolute terms. Among the three outputs, \(y^1\) (voltage) consistently exhibits the smallest RMSE, while \(y^2\) (active power) and \(y^3\) (reactive power) show similar error magnitudes. For scaling factors within the physically relevant range, the RMSE remains on the order of \(10^{-2}\,\si{pu}\), which is considered acceptable for the application under study. These results indicate that, from the perspective of the VSC-HVDC interface, the system exhibits sufficiently linear behavior to justify the use of LTI-based data-driven predictive control methods.

\section{Model-Based Industry-Standard Control}
\label{sec:lead_lag}

In this paper, we use two model-based controllers: Modal Residue (MR) and Loop-shaping (LS).

MR method \cite{gibbard2015} is a well-established approach for designing lead–lag compensators aimed at enhancing system damping.
After linearizing the system dynamics using an accurate system model, 
we designed an MR-based compensator to increase the damping of the critical \SI{0.44}{\hertz} mode from approximately \SI{0.2}{\percent} to \SI{15}{\percent} based on an I/O pairing of $u^3$ and $y^2$.
Furthermore, attempts were made to design further lead-lag controllers to damp the two remaining \SI{0.77}{\hertz} inter-area modes by pairing the other inputs and outputs.
However, these attempts were unsuccessful as they degraded the damping of  the dominant \SI{0.44}{\hertz} mode, and consequently only one of the control inputs is used by the MR design. 
This highlights a downside of single-input, single-output (SISO) control, and we note that the application of the standard MR approach was only partially successful due to the MIMO nature of our system.
 
We also implemented a MIMO LS controller using Matlab’s \texttt{loopsyn} command with the goal of rejecting disturbances with bandwidth over \SI{1}{\radian\per\second}. To focus the damping action on the desired frequency range, the plant model was augmented with an $8^\mathrm{th}$-order Butterworth band-pass filter, with cutoff frequencies of \SI{0.35}{\hertz} and \SI{2}{\hertz}.

Intended as benchmark designs for the data-driven controllers, the MR method accurately represents current industry state-of-the-art in damping controller design, while the LS represents a standard systematic approach well-known to control system practitioners.



\section{Data-driven Predictive Control (DPC)}
\label{sec:DPC}
DPC consists of two stages:
\begin{itemize}
    \item Offline: The system is excited to generate an I/O trajectory rich enough to capture the system dynamics. In our experiments, excitation is achieved using a zero-mean Gaussian white noise signal with a standard deviation of \SI{0.0025}{pu}, bounded to a maximum amplitude of \SI{0.1}{pu} to avoid interfering with the overall system operation. 
    
    \item Online: At each sampling instant, the controller solves a finite-horizon Optimal Control Problem (OCP) using the most recent $\tau_\mathrm{p}$ (past) measurements to compute an optimal sequence of control inputs over a prediction horizon $\tau_\mathrm{f}$ (future). Following the receding horizon principle, only the first input is applied to the system, and the process is repeated at the next time step.
\end{itemize}

\subsection{Data-enabled Predictive Control (DeePC)}
The DeePC algorithm \cite{coulson2019dataenabledpredictivecontrolshallows} is grounded in Willems’ Fundamental Lemma \cite{willems2005}, which states that all possible trajectories of a LTI system can be represented as linear combinations of segments from a single persistently excited trajectory.

\subsubsection*{Offline}
The I/O trajectory collected during excitation is segmented into overlapping subtrajectories of length $\tau_\mathrm{p} + \tau_\mathrm{f}$ and arranged into Hankel matrices:
\begin{align*}
    U_\mathrm{p} &= \mathcal{H}(\textbf{u}_\mathrm{data})_{[0, \tau_\mathrm{p}-1], n_\mathrm{col}}, \\
    U_\mathrm{f} &= \mathcal{H}(\textbf{u}_\mathrm{data})_{[\tau_\mathrm{p}, \tau_\mathrm{p} + \tau_\mathrm{f}], n_\mathrm{col}}, \\
    Y_\mathrm{p} &= \mathcal{H}(\textbf{y}_\mathrm{data})_{[0, \tau_\mathrm{p}-1], n_\mathrm{col}}, \\
    Y_\mathrm{f} &= \mathcal{H}(\textbf{y}_\mathrm{data})_{[\tau_\mathrm{p}, \tau_\mathrm{p} + \tau_\mathrm{f}], n_\mathrm{col}},
\end{align*}
where $n_\mathrm{col} = n_\mathrm{samples} - (\tau_\mathrm{p} + \tau_\mathrm{f}) + 1$.

\subsubsection*{Online}
At runtime, DeePC solves the following convex quadratic optimization problem:
\begin{align} \label{eq:deepc}
    \min_{\textbf{g}, \pmb{\sigma}, \textbf{u} \in \mathcal{U}, \textbf{y} \in \mathcal{Y}} \quad & \| \textbf{y} \|_{Q}^2 + \| \textbf{u} \|_{R}^2 + \lambda_{g_2} \| \textbf{g} \|_2^2 + \lambda_\sigma \| \pmb{\sigma} \|_2^2 \notag \\
    \text{s.t.} \quad &
    \begin{bmatrix}
        U_\mathrm{p} \\
        Y_\mathrm{p} \\
        U_\mathrm{f} \\
        Y_\mathrm{f} \\
    \end{bmatrix} \textbf{g} =
    \begin{bmatrix}
        \textbf{u}_\mathrm{p} \\
        \textbf{y}_\mathrm{p} + \pmb{\sigma} \\
        \textbf{u} \\
        \textbf{y} \\
    \end{bmatrix}. \tag{DeePC}
\end{align}
The optimization seeks a coefficient vector $\textbf{g}$ that generates I/O trajectories consistent with the collected data and minimizes a quadratic cost on the control effort and output deviation, here relative to a reference equal to zero, as we subtracted the mean from the real signal. Regularization on $\textbf{g}$ encourages solutions that match the true system physics, while the slack variable $\pmb{\sigma}$ ensures feasibility despite noise in the $\textbf{u}_\mathrm{p}$ and $\textbf{y}_\mathrm{p}$ measurements \cite{coulson2021distributionally, huang2021quadraticregularizationdeepc, dörfler2021bridgingdirectindirect}.
We consider the quadratic DeePC formulation, recognizing that the literature proposes other cost function formulations as well, such as L1-norm penalization \cite{coulson2019dataenabledpredictivecontrolshallows, coulson2021distributionally, cummins2025deepcHunt}.

The weighting matrices $Q$ and $R$ are block-diagonal and penalize future outputs and control inputs across the prediction horizon $\tau_\mathrm{f}$:
\begin{align*}
    Q &= \bar{Q} \otimes I \in \mathbb{R}_{\succeq 0}^{p  \tau_\mathrm{f} \times p  \tau_\mathrm{f}}, \\
    R &= \bar{R} \otimes I \in \mathbb{R}_{\succ 0}^{m  \tau_\mathrm{f} \times m  \tau_\mathrm{f}},
\end{align*}
where $\otimes$ is the Kronecker product, $\bar{Q} \in \mathbb{R}^{p \times p} \succeq 0 $ and $\bar{R} \in \mathbb{R}^{m \times m} \succeq 0$ are the cost matrices for the outputs and inputs respectively at a single timestep.

\subsection{Transient Predictive Control (TPC)}
TPC belongs to a different class of data-driven control methods. While DeePC relies on LTI subspace-based trajectory optimization, the Transient Predictor \cite{2024tpc} is constructed from LTI ARX models. Its closed-form predictor allows for fast and scalable online control when compared with DeePC, which optimizes over $\textbf{g}$. 

\subsubsection*{Offline}
Similar to other DPC methods, persistently excited I/O data is collected from the system. This data is handled to construct the two predictor matrices, $\hat{H}_\mathrm{p}$ and $\hat{H}_u$ (see \textit{Appendix}~\ref{sec:appendix_tpc_derivation} for a detailed explanation):
\begin{equation*}
    \textbf{y} = \hat{H}_\mathrm{p} \textbf{z}_\mathrm{p} + \hat{H}_u \textbf{u},
\end{equation*}
where $\mathbf{z}_\mathrm{p}$ contains the interleaved vectors of past outputs $\mathbf{y}_\mathrm{p}$ and inputs $\mathbf{u}_\mathrm{p}$, $\mathbf{u}$ is the planned future inputs, and $\mathbf{y}$ is the predicted future outputs.

\subsubsection*{Online}
At each time step, the controller solves a convex quadratic optimization problem:
\label{eq:optimization_problem_tp}
\begin{align} \label{eq:tpc}
    \min_{\textbf{u} \in \mathcal{U}, \textbf{y} \in \mathcal{Y}} \quad & \| \textbf{y} \|_{Q}^2 + \| \textbf{u} \|_{R}^2 \notag \\
    \text{s.t.} \quad & \textbf{y} = \hat{H}_\mathrm{p} \textbf{z}_\mathrm{p} + \hat{H}_u \textbf{u}. \tag{TPC}
\end{align}
As with DeePC, the weight matrices $Q$ and $R$ encode control effort and performance objectives over the horizon. Regularization can be incorporated \cite{chiuso2025harnessing, breschi2023} to address the uncertainty in the prediction, but is omitted in this work.

\subsection{Single-ARX Controller} \label{sec:single_arx}
The Single-ARX Controller, elsewhere referred to as the ``Fixed-Length Predictor'' \cite{2024tpc}, is similar to TPC but uses a single fixed-length ARX model to plan into the future rather than increasing-length ARX models. 

\subsubsection*{Offline}
As in other DPC methods, persistently excited I/O data is collected from the system. 
The data is used to construct a single $\tau_\mathrm{p}$-long ARX model which is replicated $\tau_\mathrm{f}$ times to plan $\tau_\mathrm{f}$ steps into the future (see \textit{Appendix}~\ref{sec:appendix_single_arx}). 

\subsubsection*{Online}
The online controller solves the same OCP as in \eqref{eq:tpc}, with the difference being in how the predictor matrices $\hat{H}_\mathrm{p}$ and $\hat{H}_u$ are estimated. 

\section{Simulation Results}
\label{sec:results}
All benchmarks and computations are performed on an HP EliteBook 845 G8 notebook equipped with an AMD Ryzen 3 PRO 5450U processor (4 cores, 8 threads), 16 GB of DDR4-3200 RAM, an integrated AMD Radeon GPU, and a 256 GB NVMe SSD, running on Windows 11 Enterprise 64-bit operating system.

\subsection{DPC Parameters}
\subsubsection*{ARX-based Controllers}
First, we select the prediction horizon. Since the fastest mode of the system has a frequency of approximately \SI{0.4}{\hertz} (i.e. a period of about \SI{2.5}{\second}), we aim to predict at least two full cycles. This motivates a prediction horizon of \SI{6}{\second}, corresponding to $\tau_\mathrm{f} = 60$.

The past horizon $\tau_\mathrm{p}$ on the other hand cannot be tuned based on system knowledge directly as the optimal choice of $\tau_\mathrm{p}$ relies on the quality of the training data \cite{chiuso2025harnessing}. We selected $\tau_\mathrm{p} = 30$ after running multiple experiments and evaluating predictor performance. 
A full list of TPC and the Single-ARX tuning parameters is provided in \autoref{table:paramsTP}.

\begin{figure}[t!]
    \centering
    \includegraphics[width=0.9\linewidth]{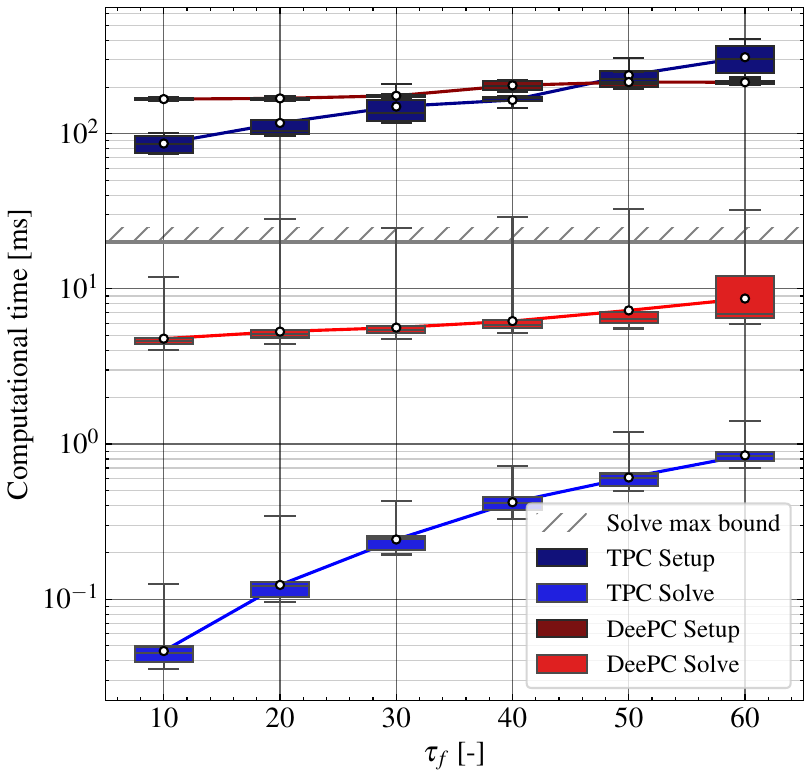}
    \caption{Setup and solve computational times using OSQP solver for TPC with $\tau_\mathrm{p} = 30$, $n_\mathrm{samples} = 2000$, and DeePC with $\tau_\mathrm{p} = 60$, $n_\mathrm{samples} = 500$.}
    \label{fig:results_comp_perf}
\end{figure}

\begin{table}[h!]
\caption{Parameters of TPC and Single-ARX.}
\label{table:paramsTP}
\centering
\begin{tabular}{l c S[table-format=4.0, table-number-alignment=center]}
\toprule
    Parameter & Symbol & $\text{Value}$ \\
\midrule
\arrayrulecolor{gray!10}
    Prediction horizon length & $\tau_\mathrm{f}$ & 60 \\
    Past horizon length & $\tau_\mathrm{p}$ & 30 \\
    Training data length & $n_\mathrm{samples}$ & 2000 \\
\midrule
    Input weights & $R$ & \text{diag$([1, \; 1, \; 1])$} \\
    Output weights & $Q$ & \text{diag$([1, \; 1, \; 1])$} \\
    Input normalization & $R_\mathrm{norm}$ & \text{diag$([0.0025, \; 0.0025, \; 0.0025])$} \\
    Output normalization & $Q_\mathrm{norm}$ & \text{diag$([0.002, \; 0.04, \; 0.02])$} \\
\arrayrulecolor{black}
\bottomrule
\end{tabular}
\end{table}

\subsubsection*{DeePC}
DeePC involves several hyperparameters and tuning parameters and is particularly sensitive to the choice of the regularization weight $\lambda_{g_2}$. We adopt the values of $\tau_\mathrm{f}$, $\tau_\mathrm{p}$, and $n_\mathrm{samples}$ from \cite{2023felix}, while the regularization weights are selected based on empirical tuning.
The final configuration is reported in \autoref{table:paramsDeePC}.

\begin{table}[h!]
\caption{Parameters of the DeePC.}
\label{table:paramsDeePC}
\centering
\begin{tabular}{l c c c}
\toprule
    Parameter & Symbol & $\text{Value}$ \\
\midrule
\arrayrulecolor{gray!10}
    Prediction horizon length & $\tau_\mathrm{f}$ & 60 \\
    Past horizon length & $\tau_\mathrm{p}$ & 60 \\
    Training data length & $n_\mathrm{samples}$ & 500 \\
\midrule
    Regularization weight & $\lambda_{g_2}$ & 30 \\
    Relaxation weight & $\lambda_\sigma$ & $10^{5}$ \\
    Input weights & $R$ & \text{diag$([10^{-2}, \; 1, \; 10^{-2}])$} \\
    Output weights & $Q$ & \text{diag$([10^{8}, \; 10^{7}, \; 10^{7}])$} \\
\arrayrulecolor{black}
\bottomrule
\end{tabular}
\end{table}

\subsection{Computational Cost}
The computational cost of DPC algorithms is important as the online OCP is run every timestep. With the understanding that the online solve time for TPC and Single-ARX are equivalent, we omit Single-ARX in this section.

The computational cost of DPC is largely dictated by the number of decision variables. For TPC, this number grows with the product of the number of inputs and the prediction horizon, i.e. $m  \tau_\mathrm{f}$. In contrast, for DeePC the number of decision variables is dominated by $\mathbf{g}$, which increases linearly with the number of samples $n_\mathrm{samples}$. This makes TPC and Single-ARX better-suited for online applications that require large data sets and fast online optimization, such as the power oscillation damping application that we investigate.

\begin{figure*}[t!]
    \centering
    \includegraphics[width=1.0\linewidth]{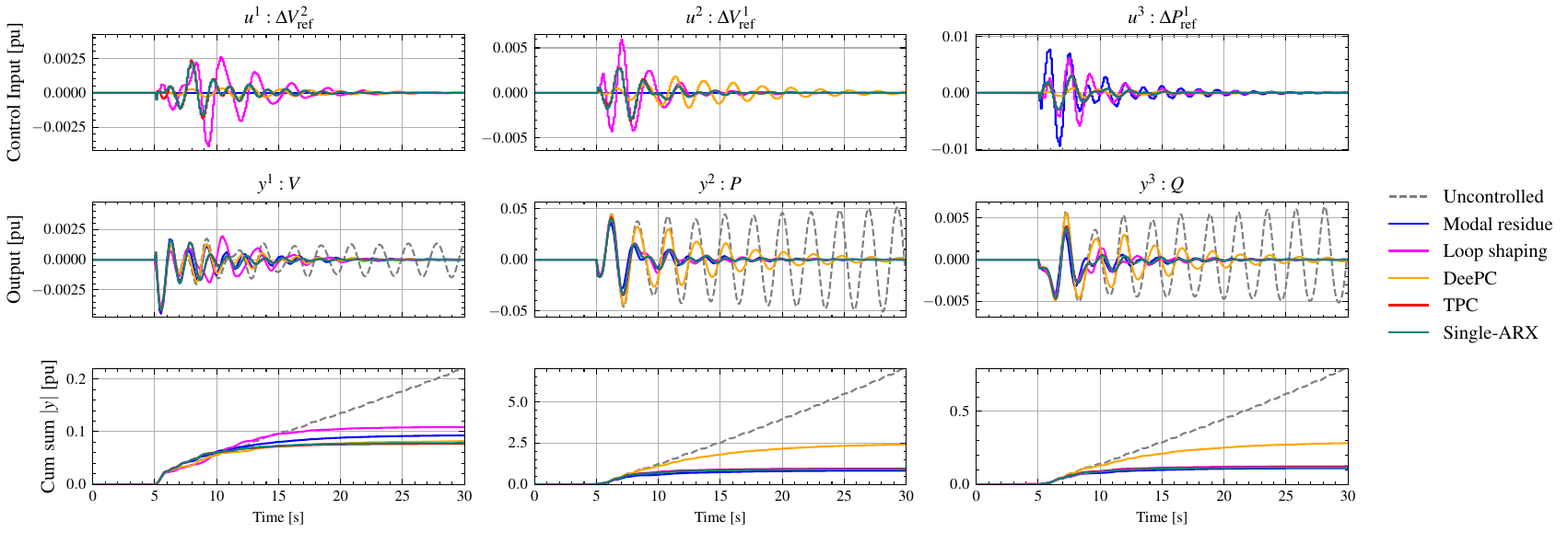}
    \caption{Control performance for tripping of transmission line 16-17 (Scenario~4) at \SI{5}{\second} with \SI{50}{\percent} of PSS deactivated. \\   
    The plots for the closely-related TPC and Single-ARX methods overlap and thus the red line is underneath the green line.}
    \label{fig:results_control_perf_pss2_scenario3}
\end{figure*}

\autoref{fig:results_comp_perf} shows the statistics for setup and solve times for both TPC and DeePC under various configurations. For power oscillation damping, the solve time is the primary constraint for real-time implementation and must remain below \SI{20}{\milli\second}. Conversely, setup is performed only once at controller initialization and thus its timing is less critical. Across the considered range of $\tau_\mathrm{f}$, TPC consistently achieves solve times well within the computational budget, and it is at least one order of magnitude faster than DeePC. The latter shows computational time dangerously close to the threshold and occasionally surpasses it.

We evaluated four open-source convex optimization solvers: Clarabel \cite{clarabel}, PIQP \cite{piqp}, SCS \cite{scs2021}, and OSQP \cite{osqp} (see \textit{Appendix}~\ref{sec:appendix_solver}). Among them, OSQP demonstrated the best performance, with the lowest average solve time and smallest variability. Notably, OSQP’s solve time approaches that of the unconstrained formulation, where the optimal control problem admits a closed-form solution. 
We attribute this to OSQP being run in parametric/warm-started mode via its \texttt{update()} interface, which reuses the previously factorized Karush–Kuhn–Tucker (KKT) system and the previous primal/dual iterates.
Since the OCP is a convex QP, global optimality is guaranteed.

\subsection{Control Performance}
The fault scenarios presented in \autoref{tab:scenarios} are use to evaluate the performance of the controllers. All the results presented in this section refer to the most challenging case that the controllers are able to stabilize---\SI{50}{\percent} of PSS deactivated. 
All easier cases ($<\SI{50}{\percent}$ of PSSs deactivated) are stabilized with equal or better performance. 
For more severe cases with \SI{75}{\percent} PSSs deactivated, all controllers, including the industry standard, fail to ensure stability. Stabilizing such cases is the subject of future work.

\begin{table}[h!]
    \caption{Different fault scenarios used for control performance evaluation (in blue in \autoref{fig:power_system}).}
    \label{tab:scenarios}
    \centering
    \begin{tabular}{c l}
        \toprule
        \textbf{Scenario} & \textbf{Description} \\ 
        \midrule
        1 & Generator 5 setpoint change \\
        2 & Generator 1 trip \\
        3 & Generator 4 trip \\
        4 & Line 16-17 trip \\
        5 & Line 6-7 trip \\
        6 & Tie-line trip \\ 
        \bottomrule
    \end{tabular}
\end{table}

\autoref{fig:results_control_perf_pss2_scenario3} shows the performance of all controllers on fault Scenario~4, when subjected to the tripping of line 16-17 at \SI{5}{\second}, simulated without noise. Without wide-area control, the system becomes unstable, whereas all tested controllers successfully dampen the oscillations, restoring stability.

For ease of comparison, the third row of the figure presents the cumulative sum of the absolute output values, used here as a proxy for the cost function. TPC and Single-ARX controllers deliver performance similar to the model-based, industry-standard options MR and LS, while DeePC provides less damping. The results are sensitive to the selected tuning parameters and thus this should not be taken as a generalizable insight. 
Both the model-based MR and LS controllers produce aggressive responses, with large control inputs. TPC and Single-ARX achieve similar performance with less control effort. 

\begin{figure}[t!]
    \centering
    \includegraphics[width=1.0\linewidth]{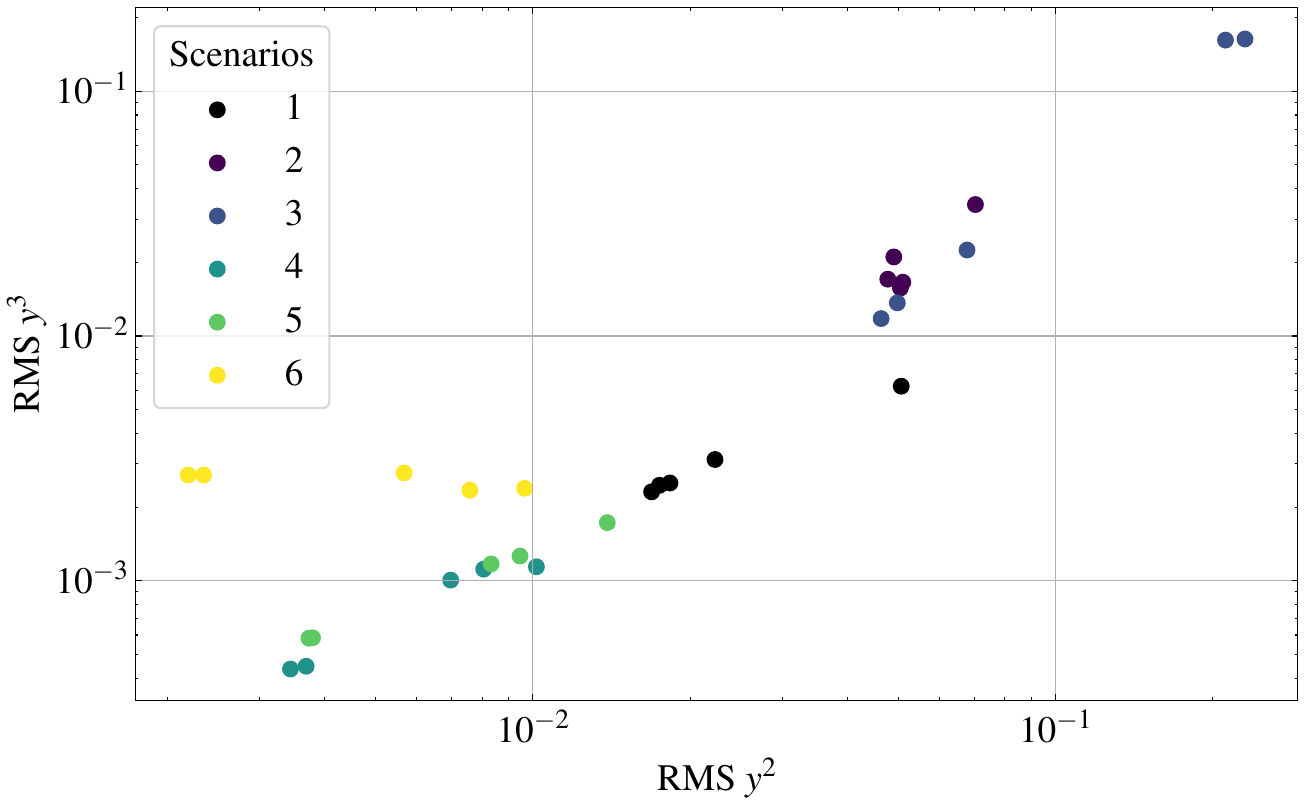}
    \caption{Correlation of the RMS of $y^2$ (active power) and the RMS of $y^3$ (reactive power) for the  simulations with noise.}
    \label{fig:results_corr}
\end{figure}

\begin{figure}[t!]
    \centering
    \includegraphics[width=1.0\linewidth]{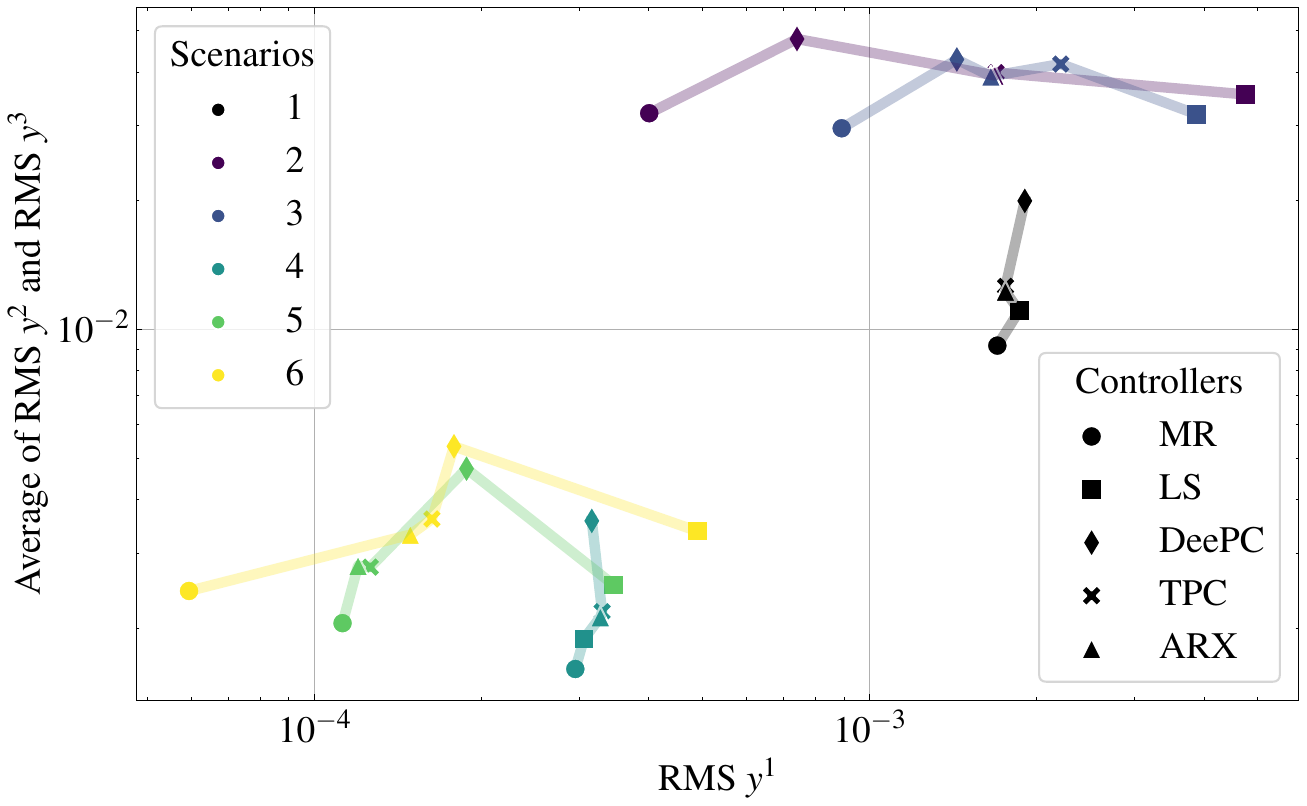}
    \caption{Performance comparison of all controllers for all scenarios in the noiseless simulations.}
    \label{fig:results_nonoise}
\end{figure}

\begin{figure}[t!]
    \centering
    \includegraphics[width=1.0\linewidth]{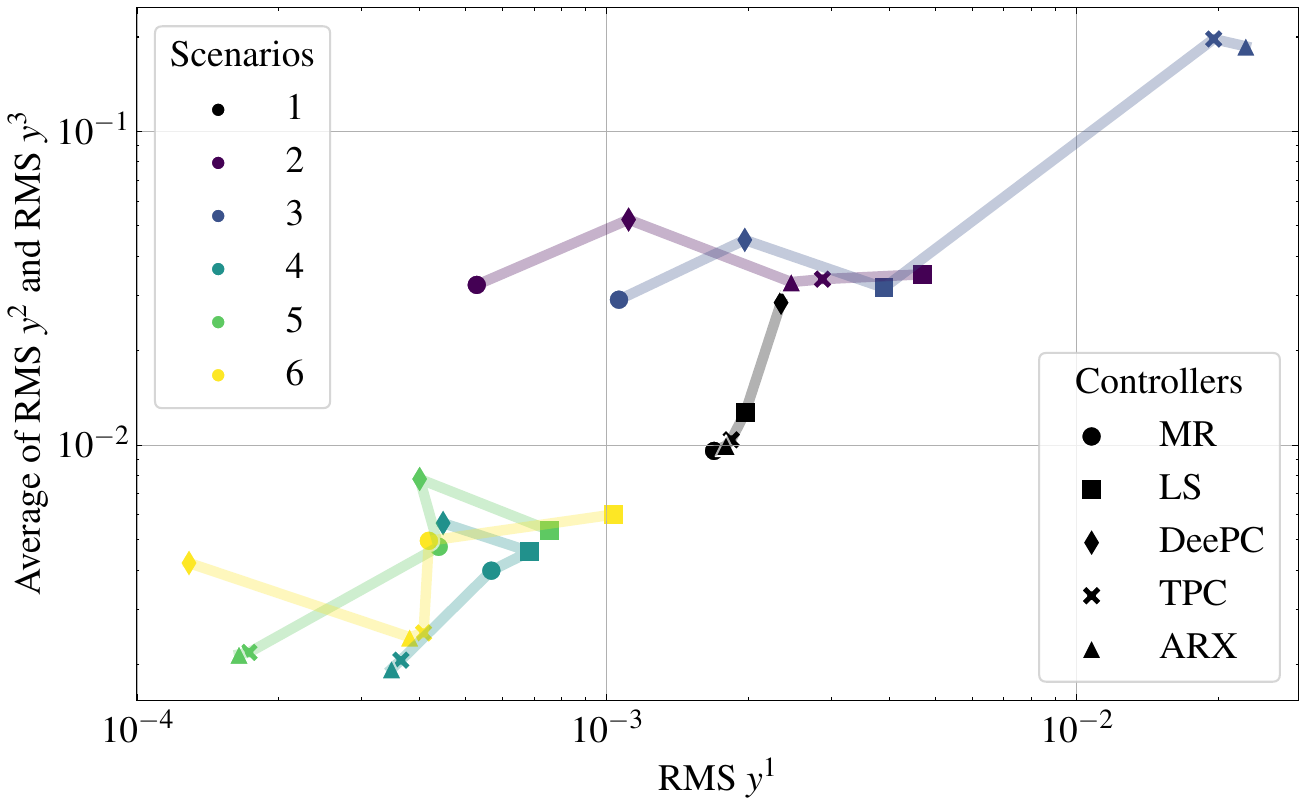}
    \caption{Performance comparison of all controllers for all scenarios in the simulations with process noise.}
    \label{fig:results_noise}
\end{figure}

The performance in terms of Root Mean Square (RMS) for all fault scenarios is shown in \autoref{fig:results_nonoise} for the noiseless case and \autoref{fig:results_noise} for the case where load variation (process noise) is present in both the training data collection and the evaluation simulations. To display the 3 outputs $y^1$, $y^2$ and $y^3$ in a visually comprehensive manner in Fig.s~\ref{fig:results_nonoise} and \ref{fig:results_noise}, we combine the $y^2$ and $y^3$ performance onto a single axis. This is motivated by \autoref{fig:results_corr}, which demonstrates that the $y^2$ (active power) and $y^3$ (reactive power) performances are strongly correlated for all scenarios.

For the scenarios without noise (\autoref{fig:results_nonoise}), we observe that the MR controller always dominates the other controllers in both objectives. TPC and Single-ARX (ARX for short in the figure) are always very close together, with Single-ARX tending to perform slightly better. We attribute the relative performance of Single-ARX to TPC to the large prediction horizon: the variance of the ARX models predicted for the large prediction lengths can be large and dominate the TPC merits, which arise when there are few training data. 
DeePC outperforms Single-ARX and TPC in scenarios 2 and 3. Once again, the relative performance of the DPC methods is sensitive to the tuning parameters and thus should not be considered to apply generally.
The primary take-away from the noiseless simulations is that the data-driven controllers are competitive with the model-based controllers.

A similar situation is visible for the simulations with process noise (\autoref{fig:results_noise}). Here, for one third of the scenarios (the line trips), ARX dominates the other controllers, while for one scenario (one generator trip) ARX and TPC are dominated. This motivates future work in which the ARX and TPC OCPs are augmented with regularization, which may help address high-variance predictions when the training data includes process noise.

\section{Conclusions}
\label{sec:conclusion}
This paper compares data-driven and conventional model-based approaches for controlling a VSC-HVDC power line to dampen inter-area power oscillations. 
In the test cases investigated (outlined in \autoref{fig:power_system} and \autoref{tab:scenarios}), all three DPC strategies perform comparably to the model-based methods in stabilizing challenging fault scenarios. Since DPC does not need system models and comes with reduced tuning effort, these approaches pave the way to oscillation damping in future grids featuring higher complexities and increased uncertainties, where system models will be even harder to obtain than today.

Our linearity analysis demonstrated that, from the perspective of the VSC-HVDC actuators, the grid dynamics are close-to-linear, which further supports linear DPC approaches for this application.
The computational costs of the DPC methods were  investigated, demonstrating that ARX-based DPC, including TPC, is less computationally expensive than DeePC and comfortably below real-time thresholds.
The results in this paper warrant further investigation into data-driven power system stabilization.




\section*{Acknowledgment}
Contributions and support from Sebastian Graf, Linbin Huang, Jeremy Coulson, and Alberto Bolzoni are gratefully acknowledged.

\bibliographystyle{IEEEtran}
\bibliography{Bibliography}

\appendices

\section{Derivation of the Transient Predictor}
\label{sec:appendix_tpc_derivation}
We consider a discrete-time LTI system where input $\mathbf{u}_t \in \mathbb{R}^m$ and output $\mathbf{y}_t \in \mathbb{R}^p$ form a joint stationary process:
\begin{equation}\label{eqn:infiniteLength}
    \textbf{z}_t =
    \begin{bmatrix}
    \textbf{y}_t \\ \textbf{u}_t
    \end{bmatrix} \in \mathbb{R}^{p+m}.
\end{equation}
The system admits an infinite-horizon ARX representation:
\begin{equation*}
    \textbf{y}_{t+1} = \sum_{i=1}^{\infty} \Phi_{\infty,i} \textbf{z}_{t-i+1} + e_{t+1},
\end{equation*}
with $\Phi_{\infty,i} \in \mathbb{R}^{p \times (p+m)}$, and $e$ is a zero-mean white noise, which results from prediction errors. Since only a finite data window can be used in practice, we estimate a truncated ARX model of length $\tau$:
\begin{equation}\label{eqn:finiteLength}
    \textbf{y}_{t+1} = \sum_{i=1}^{\tau} \phi_{\tau,i} \textbf{z}_{t-i+1} + \epsilon_{\tau,t+1},
\end{equation}
where $\phi_{\tau,i} \in \mathbb{R}^{p \times (p+m)}$ is the $i^{\text{th}}$ entry of the $\tau$-long optimal (minimum variance, linear, unbiased) single-step predictor, and $\epsilon_{\tau, t+1}$ is the transient innovation due to limited lead-in data and unknown initial conditions.

\subsubsection*{Offline}
From the collected excitation data, the I/O trajectories are interleaved with $\textbf{y}_\mathrm{data}$ for each timestep put on top of $\textbf{u}_\mathrm{data}$ for the same timestep to form $\textbf{z}_\mathrm{data}$.
$\textbf{z}_\mathrm{data}$ is then segmented into overlapping subtrajectories of length $\tau_\mathrm{p} + \tau_\mathrm{f}$ to form a row-interleaved Hankel matrix
\begin{equation*}
    Z = \mathcal{H}(\textbf{z}_\mathrm{data})_{[0, \tau_\mathrm{p} + \tau_\mathrm{f}], n_\mathrm{col}},
\end{equation*}
where $n_\mathrm{col} = n_\mathrm{samples} - (\tau_\mathrm{p} + \tau_\mathrm{f}) + 1$.

The offline calculations seek to find increasing-length ARX model parameters that fit
$
    Y_\mathrm{f}^\uparrow = \Phi  Z,
$
where $Y_\mathrm{f}^\uparrow = \mathcal{H}(\textbf{z}_\mathrm{data})_{[\tau_\mathrm{p}+1, \tau_\mathrm{p} + \tau_\mathrm{f}], n_\mathrm{col}}$ is the Hankel matrix of output values starting with time $\tau_\mathrm{p}+1$ as the upper-left entry.
As the data is noisy and the system is not LTI, there is no $\Phi$ that satisfies $Y_\mathrm{f}^\uparrow = \Phi  Z$. Thus, we estimate a best-fit $\hat{\Phi}$.

LQ-decomposing $Z$, we get $Z$ expressed as the linear combination of the orthogonal rows of $\mathrm{Q}$, 
\begin{equation*}
    L \mathrm{Q} = \texttt{LQ}(Z),
\end{equation*}
where $L$ is lower triangular and $\mathrm{Q}$ is orthogonal. 
To estimate $\hat{\Phi}$, we construct the following equation
\begin{equation} \label{eq:arx_lq_base}
    L_y^0 \mathrm{Q} = \hat{\Phi}  L \mathrm{Q},
\end{equation}
where $L_y^0$ is obtained by
\begin{enumerate}
    \item removing the rows of $L$ corresponding to inputs,
    \item removing the rows of $L$ corresponding to past measurements (the first $p  \tau_\mathrm{p}$ rows), and
    \item eliminating block entries along the diagonal. These entries of $L$ describe the portion of $Y_\mathrm{f}^\uparrow$ that is not a linear combination of previous measurements.
\end{enumerate}
Thus, $L_y^0 \mathrm{Q}$ is the portion of $Y_\mathrm{f}^\uparrow$ that \textit{is} a linear combination of previous measurements.

\begin{figure*}[!t]
    \begin{alignat}{4}
        \hat{\redY}_{t+1} &= \hat{\hat{\Phi}}^{\redy}_{\tau_\mathrm{p},\tau_\mathrm{p}} \redY_{t-\tau_\mathrm{p}+1} &&+ \dots + \hat{\hat{\Phi}}^{\redy}_{\tau_\mathrm{p},1} \redY_{t} && && \notag \\
        &+ \hat{\hat{\Phi}}^{\blueu}_{\tau_\mathrm{p},\tau_\mathrm{p}} \blueU_{t-\tau_\mathrm{p}+1} &&+ \dots + \hat{\hat{\Phi}}^{\blueu}_{\tau_\mathrm{p},1} \blueU_{t} && && \notag \\
        \hat{\redY}_{t+2} &= \hat{\hat{\Phi}}^{\redy}_{\tau_\mathrm{p}+1,\tau_\mathrm{p}+1} \redY_{t-\tau_\mathrm{p}+1} &&+ \dots + \hat{\hat{\Phi}}^{\redy}_{\tau_\mathrm{p}+1,2} \redY_{t} &&+ \hat{\hat{\Phi}}^{\redy}_{\tau_\mathrm{p}+1,1} \hat{\redY}_{t+1} && \notag \\
        &+ \hat{\hat{\Phi}}^{\blueu}_{\tau_\mathrm{p}+1,\tau_\mathrm{p}+1} \blueU_{t-\tau_\mathrm{p}+1} &&+ \dots + \hat{\hat{\Phi}}^{\blueu}_{\tau_\mathrm{p}+1,2} \blueU_{t} &&+ \hat{\hat{\Phi}}^{\blueu}_{\tau_\mathrm{p}+1,1} \blueU_{t+1} && \label{eqn:fullTPC} \\
        \vdots \;\; &= \qquad \vdots &&+ \dots + \quad \vdots &&+ \quad \vdots &&+ \ddots \notag \\
        \hat{\redY}_{t+\tau_\mathrm{f}} &= \hat{\hat{\Phi}}^{\redy}_{\tau_\mathrm{p}+\tau_\mathrm{f},\tau_\mathrm{p}+\tau_\mathrm{f}} \redY_{t-\tau_\mathrm{p}+1} &&+ \dots + \hat{\hat{\Phi}}^{\redy}_{\tau_\mathrm{p}+\tau_\mathrm{f},\tau_\mathrm{f}+1} \redY_{t} &&+ \hat{\hat{\Phi}}^{\redy}_{\tau_\mathrm{p}+\tau_\mathrm{f},\tau_\mathrm{f}} \hat{\redY}_{t+1} &&+ \dots + \hat{\hat{\Phi}}^{\redy}_{\tau_\mathrm{p}+\tau_\mathrm{f},1} \hat{\redY}_{t+\tau_\mathrm{f}-1} \notag \\
        \underbrace{\phantom{\hat{\redY}_{t+\tau_\mathrm{f}}}}_{\displaystyle\hat{\textbf{y}}} &+ \tikzmarknode{start_p}{\hat{\hat{\Phi}}^{\blueu}_{\tau_\mathrm{p}+\tau_\mathrm{f},\tau_\mathrm{p}+\tau_\mathrm{f}} \blueU_{t-\tau_\mathrm{p}+1}} &&+ \dots + \tikzmarknode{end_p}{\hat{\hat{\Phi}}^{\blueu}_{\tau_\mathrm{p}+\tau_\mathrm{f},\tau_\mathrm{f}+1} \blueU_{t}} &&+ \tikzmarknode{start_f}{\hat{\hat{\Phi}}^{\blueu}_{\tau_\mathrm{p}+\tau_\mathrm{f},\tau_\mathrm{f}} \blueU_{t+1}} &&+ \dots + \hat{\hat{\Phi}}^{\blueu}_{\tau_\mathrm{p}+\tau_\mathrm{f},1} \tikzmarknode{end_f}{\blueU_{t+\tau_\mathrm{f}-1}} \notag
    \end{alignat}
    \begin{tikzpicture}[overlay, remember picture]
        \draw[decorate, decoration={brace, amplitude=5pt, mirror}, thick] 
            (start_p.south west) -- (end_p.south east) 
            node[midway, below=6pt] {$\hat{\Phi}_\mathrm{p} \textbf{z}_\mathrm{p}$};
    \end{tikzpicture}
    \begin{tikzpicture}[overlay, remember picture]
        \draw[decorate, decoration={brace, amplitude=5pt, mirror}, thick] 
            (start_f.south west) -- (end_f.south east) 
            node[midway, below=6pt] {$\hat{\Phi}_y \textbf{y} + \hat{\Phi}_u \textbf{u}$};
    \end{tikzpicture}
\end{figure*}

Right-multiplying \eqref{eq:arx_lq_base} by $\mathrm{Q}^{\top}$, leveraging that $Q$ is orthogonal and thus $Q Q^\top = I$, gives $L_y^0 = \hat{\Phi} L$. Right-multiplying this equation by $L^{-1}$ yields the increasing-length ARX predictor matrix:
\begin{align*}
    \hat{\Phi} &= L_y^0 L^{-1} \\
    &=
    \begin{bmatrix}
         \quad \horzbar & \hat{\Phi}_1 & \horzbar & 0 & & \hdots & 0 \\
         \quad \horzbar & \qquad \hat{\Phi}_2 & \qquad \horzbar & & 0 & \hdots & 0 \\
         &   &  \ddots \qquad \qquad & & & \ddots & \vdots \\
         \quad \horzbar & & \quad \hat{\Phi}_\tau & & \horzbar & & 0 
    \end{bmatrix} . \\ 
\end{align*}
Asserting new definitions for the submatrices of $\hat{\Phi}$, we get
\begin{align*}
    \hat{\Phi} &=  \left[\hat{\Phi}_\mathrm{p} \quad \hat{\Phi}_{y,1} \quad \hat{\Phi}_{u,1} \quad \dots \quad \hat{\Phi}_{y,\tau_\mathrm{f}} \quad \hat{\Phi}_{u,\tau_\mathrm{f}} \right],
\end{align*}
where $\hat{\Phi}_\mathrm{p}$ is a dense matrix that is right-multiplied by past measurements.
We can now rearrange the parameters that are right-multiplied by future values based on whether the future values are outputs or inputs:
\begin{equation*}
    \hat{\Phi}_y = 
    \begin{bmatrix}
        | & | & & | \\
        \hat{\Phi}_{y,1} & \hat{\Phi}_{y,2} & \dots & \hat{\Phi}_{y,\tau_\mathrm{f}}\\
        | & | & & | \\
    \end{bmatrix} = \tikztriangleLow,
\end{equation*}
\begin{equation*}
    \hat{\Phi}_u = 
    \begin{bmatrix}
        | & | & & | \\
        \hat{\Phi}_{u,1} & \hat{\Phi}_{u,2} & \dots & \hat{\Phi}_{u,\tau_\mathrm{f}} \\
        | & | & & | \\
    \end{bmatrix} = \tikztriangleLow.
\end{equation*}
The triangles indicate the lower-triangular structure of these matrices.
The resulting multi-step predictor gives the TPC output prediction \cite{2024tpc}:
\begin{equation} \label{eq:tpc_prediction_eq}
    \hat{\textbf{y}} = \hat{\Phi}_\mathrm{p}
    \begin{bmatrix} 
        \textbf{y}_\mathrm{p} \\ \textbf{u}_\mathrm{p} 
    \end{bmatrix} 
    + \hat{\Phi}_y \hat{\textbf{y}} + \hat{\Phi}_u \textbf{u}. 
\end{equation}
The TPC output prediction is described in detail in \eqref{eqn:fullTPC} using the definition for $\phi_{\tau,i}$ in \eqref{eqn:finiteLength}. Solving \eqref{eq:tpc_prediction_eq} for $\hat{\mathbf{y}}$ gives
\begin{equation*}
    \hat{\textbf{y}} = \hat{H}_\mathrm{p} \begin{bmatrix} 
    \textbf{y}_\mathrm{p} \\ \textbf{u}_\mathrm{p} 
    \end{bmatrix} 
    + \hat{H}_u \textbf{u}. 
\end{equation*}
Therefore, the TPC consists of the following matrices:
\begin{align*}
    \hat{H}_\mathrm{p} &= (I - \hat{\Phi}_y)^{-1} \hat{\Phi}_\mathrm{p} = \tikzsquare \in \mathbb{R}^{p  \tau_\mathrm{f} \times (p + m)\tau_\mathrm{p}}, \\
    \hat{H}_u &= (I - \hat{\Phi}_y)^{-1} \hat{\Phi}_u = \tikztriangleLow \in \mathbb{R}^{p  \tau_\mathrm{f} \times m  \tau_\mathrm{f}}.
\end{align*}
Importantly, the strict lower-triangular structure of $\hat{H}_u$ ensures strict causality of the TPC. 
An open-source code base that constructs the TPC can be found at \url{https://github.com/do-dpc/do-dpc?tab=readme-ov-file}.

\section{Derivation of the Fixed-Length Predictor}
\label{sec:appendix_single_arx}
As in TPC, I/O data are vertically interleaved and the Hankel matrix $Z$ is constructed. Unlike TPC, we estimate a single-step ARX model of fixed length $\tau_\mathrm{p}$ and replicate it across the prediction horizon $\tau_\mathrm{f}$. This results in a repeated block structure for the multi-step predictor:
\begin{align*}
    \hat{\Psi} &=
    \begin{bmatrix}
        - & \hat{\Phi}_1 & - & 0 & \cdots & & 0 \\
        0 & - & \hat{\Phi}_1 & - & 0 & \cdots & 0 \\
        & & & \ddots & & \\
        0 & \cdots & 0 & - & \hat{\Phi}_1 & - & 0 &
    \end{bmatrix} . 
\end{align*}

\begin{figure*}[!t]
    \begin{alignat}{5}
        \hat{\redY}_{t+1} &= \hat{\phi}^{\redy}_{\tau_\mathrm{p},\tau_\mathrm{p}} \redY_{t-\tau_\mathrm{p}+1} +\; &&\hat{\phi}^{\redy}_{\tau_\mathrm{p},\tau_\mathrm{p}-1} \redY_{t-\tau_\mathrm{p}+2} &&+ \dots + \hat{\phi}^{\redy}_{\tau_\mathrm{p},1} \redY_{t} && && \notag\\
        &+ \hat{\phi}^{\blueu}_{\tau_\mathrm{p},\tau_\mathrm{p}} \blueU_{t-\tau_\mathrm{p}+1} +\; &&\hat{\phi}^{\blueu}_{\tau_\mathrm{p},\tau_\mathrm{p}-1} \blueU_{t-\tau_\mathrm{p}+2} &&+ \dots + \hat{\phi}^{\blueu}_{\tau_\mathrm{p},1} \blueU_{t} && && \notag \\
        \hat{\redY}_{t+2} &= &&\hat{\phi}^{\redy}_{\tau_\mathrm{p},\tau_\mathrm{p}} \redY_{t-\tau_\mathrm{p}+2} &&+ \dots + \hat{\phi}^{\redy}_{\tau_\mathrm{p},2} \redY_{t} &&+ \hat{\phi}^{\redy}_{\tau_\mathrm{p},1} \hat{\redY}_{t+1} && \label{eqn:fullARX} \\
        &+ \phantom{y} &&\hat{\phi}^{\blueu}_{\tau_\mathrm{p},\tau_\mathrm{p}} \blueU_{t-\tau_\mathrm{p}+2} &&+ \dots + \hat{\phi}^{\blueu}_{\tau_\mathrm{p},2} \blueU_{t} &&+ \hat{\phi}^{\blueu}_{\tau_\mathrm{p},1} \blueU_{t+1} && \notag \\
        \vdots \;\; &= && \qquad \qquad \ddots &&   \quad  && \quad  && \ddots \notag \\
        \hat{\redY}_{t+\tau_\mathrm{f}} &= && &&\phantom{+} \;  &&  \hat{\phi}^{\redy}_{\tau_\mathrm{p},\tau_\mathrm{p}} \hat{\redY}_{t+\tau_\mathrm{f}-\tau_\mathrm{p}} &&+ \dots + \hat{\phi}^{\redy}_{\tau_\mathrm{p},1} \hat{\redY}_{t+\tau_\mathrm{f}-1} \notag \\
        \underbrace{\phantom{\hat{\redY}_{t+\tau_\mathrm{f}}}}_{\displaystyle\hat{\textbf{y}}} &+ \tikzmarknode{start_p2}{} && &&\phantom{+} \;  \tikzmarknode{end_p2}{\phantom{+ \hat{\phi}^{\blueu}_{t-1} \blueU_{t}}} && \tikzmarknode{start_f2} {\hat{\phi}^{\blueu}_{\tau_\mathrm{p},\tau_\mathrm{p}} \blueU_{t+\tau_\mathrm{f}-\tau_\mathrm{p}}} &&+ \dots + \hat{\phi}^{\blueu}_{\tau_\mathrm{p},1} \tikzmarknode{end_f2}{\blueU_{t+\tau_\mathrm{f}-1}} \notag
    \end{alignat}
    \begin{tikzpicture}[overlay, remember picture]
        \draw[decorate, decoration={brace, amplitude=5pt, mirror}, thick] 
            (start_p2.south west |- end_p2.south east) -- (end_p2.south east) 
            node[midway, below=6pt] {$\hat{\Psi}_\mathrm{p} \textbf{z}_\mathrm{p}$};
    \end{tikzpicture}
    \begin{tikzpicture}[overlay, remember picture]
        \draw[decorate, decoration={brace, amplitude=5pt, mirror}, thick] 
            (start_f2.south west) -- (end_f2.south east) 
            node[midway, below=6pt] {$\hat{\Psi}_y \textbf{y} + \hat{\Psi}_u \textbf{u}$};
    \end{tikzpicture}
\end{figure*}

The Single-ARX prediction is described as follows:
\begin{equation*}\label{eq:singleARX_prediction_eq}
    \hat{\textbf{y}} = \hat{\Psi}_\mathrm{p}
    \begin{bmatrix} 
        \textbf{y}_\mathrm{p} \\ \textbf{u}_\mathrm{p} 
    \end{bmatrix} 
    + \hat{\Psi}_y \textbf{y} + \hat{\Psi}_u \textbf{u}.
\end{equation*}
\eqref{eqn:fullARX} provides a more detailed description, where the summation terms are explicitly presented. Each term in \eqref{eqn:fullARX} corresponds to a $\phi_{\tau_\mathrm{p},i}$ term in \eqref{eqn:finiteLength}, where $i$ describes the distance into the past at time $t$.

From $\hat{\Psi}$, we compute the prediction matrices $\hat{H}_\mathrm{p}$ and $\hat{H}_u$ as in TPC. Again, causality is enforced by the strictly lower-triangular shape of $\hat{H}_u$.

\section{Solver Computation Comparison}
\label{sec:appendix_solver}
The body of this paper implements the OSQP solver. Here we show how the OSQP solver compares with three other open-source solvers, such as Clarabel, SCS, and PIQP, as well as with closed-form solution for the unconstrained case for TPC and Subspace Predictive Control, which is closely related to DeePC. All solvers were implemented in Python.

\subsection{Constrained Optimal Control Formulations}\label{sec:constrainedOCPformulations}

\subsubsection*{DeePC}
The literature proposes DeePC cost functions including both 1 and 2-norm regularization \cite{dörfler2021bridgingdirectindirect}. In this work, we focus on just 2-norm regularization as shown in \eqref{eq:deepc}.
By substituting for $\mathbf{g}$ and $\boldsymbol{\sigma}$ in \eqref{eq:deepc} and using the decision variable vector
\begin{equation*}
    \mathbf{x} = \begin{bmatrix} \mathbf{g} \\ \boldsymbol{\sigma} \end{bmatrix} \in \mathbb{R}^{n_\text{col}+p  \tau_\mathrm{p} },
\end{equation*}
the OCP can be rewritten as
\begin{subequations}
\begin{align*}
    \min_{\mathbf{x}} \quad & \mathbf{x}^\top P_\mathrm{DeePC} \mathbf{x} +
    2 \textbf{q}_\mathrm{DeePC}^\top \mathbf{x} \\
    \text{s.t.} \quad &
    \begin{bmatrix}
        U_\mathrm{p} & 0 \\
        Y_\mathrm{p} & -I \\
    \end{bmatrix} \mathbf{x} = 
    \begin{bmatrix}
        \textbf{u}_\mathrm{p} \\
        \textbf{y}_\mathrm{p}
    \end{bmatrix}, \\
    & \begin{bmatrix}
        \textbf{u}_\mathrm{lb} \\
        \textbf{y}_\mathrm{lb}
    \end{bmatrix} \leq 
    \begin{bmatrix}
        U_\mathrm{f} & 0 \\
        Y_\mathrm{f} & 0 \\
    \end{bmatrix}
    \mathbf{x} \leq
    \begin{bmatrix}
        \textbf{u}_\mathrm{ub} \\
        \textbf{y}_\mathrm{ub}
    \end{bmatrix},
\end{align*}
\end{subequations}
where 
\begin{align*}
    P_\mathrm{DeePC} &= 
    \begin{bmatrix}
        Y_\mathrm{f}^\top Q Y_\mathrm{f} + U_\mathrm{f}^\top R U_\mathrm{f} + \lambda_{g_2} I & 0 \\
        0 & \lambda_\sigma I \\
    \end{bmatrix}, 
    \\
    \textbf{q}_\mathrm{DeePC} &= 0 \in \mathbb{R}^{n_\text{col}+p  \tau_\mathrm{p} }.
\end{align*}


\subsubsection*{TPC}
To reduce the size of the \eqref{eq:tpc} optimization we substitute for $\textbf{y}$, obtaining the following OCP in $\textbf{u}$:
\begin{subequations}
\begin{align*}
    \min_{\textbf{g}, \boldsymbol{\sigma}} \quad & \textbf{u}^\top P_\mathrm{TPC} \textbf{u} + 2 \textbf{q}_\mathrm{TPC}^\top \textbf{u} \\
    \text{s.t.} \quad &
    \begin{bmatrix}
        \textbf{u}_\mathrm{lb} \\
        \textbf{y}_\mathrm{lb} - \hat{H}_\mathrm{p} \textbf{z}_\mathrm{p}
    \end{bmatrix} \leq 
    \begin{bmatrix}
        I \\
        \hat{H}_u \\
    \end{bmatrix}
    \textbf{u} \leq
    \begin{bmatrix}
        \textbf{u}_\mathrm{ub} \\
        \textbf{y}_\mathrm{ub} - \hat{H}_\mathrm{p} \textbf{z}_\mathrm{p}
    \end{bmatrix},
\end{align*}
\end{subequations}
where
\begin{align*}
    P_\mathrm{TPC} &= \hat{H}_u^\top Q \hat{H}_u + R \in \mathbb{R}^{p  \tau_\mathrm{f} \times p  \tau_\mathrm{f}}, \\
     \textbf{q}_\mathrm{TPC} &= \hat{H}_u^\top Q \hat{H}_\mathrm{p} \textbf{z}_\mathrm{p}\in \mathbb{R}^{p  \tau_\mathrm{f} }.
\end{align*}

\subsection*{Quadratic Program Solvers: OSQP and PIQP}

Taking the following Quadratic Program (QP) form, 

\begin{subequations}
\begin{align*}
    \min\limits_{\textbf{x}} \quad & \frac{1}{2} \textbf{x}^\top P \textbf{x} + \textbf{q}^\top \textbf{x} \\
    \text{s.t.} \quad & A \textbf{x} = \textbf{b} \\
    & G \textbf{x} \leq \textbf{h} \\
    &  \textbf{x}_\mathrm{lb} \leq \textbf{x} \leq  \textbf{x}_\mathrm{ub},
\end{align*}
\end{subequations}
the constraints for DeePC are
\begin{align*}
    A &= 
    \begin{bmatrix}
        U_\mathrm{p} & 0 \\
        Y_\mathrm{p} & -I \\
    \end{bmatrix} \in \mathbb{R}^{(m+p)\tau_\mathrm{p} \times (n_\mathrm{col} + p  \tau_\mathrm{p})}, \\
    \textbf{b} &= 
    \begin{bmatrix}
        \textbf{u}_\mathrm{p} \\
        \textbf{y}_\mathrm{p}
    \end{bmatrix} \in \mathbb{R}^{(m+p)\tau_\mathrm{p} }, \\
    G &=
    \begin{bmatrix}
        U_\mathrm{f} & 0 \\
        Y_\mathrm{f} & 0 \\
        -U_\mathrm{f} & 0 \\
        -Y_\mathrm{f} & 0 \\
    \end{bmatrix} \in \mathbb{R}^{(m+p)2\tau_\mathrm{f} \times (n_\mathrm{col} + p  \tau_\mathrm{p})}, \\
    \textbf{h} &= 
    \begin{bmatrix}
        \textbf{u}_\mathrm{ub} \\
        \textbf{y}_\mathrm{ub} \\
        -\textbf{u}_\mathrm{lb} \\
        -\textbf{y}_\mathrm{lb} \\
    \end{bmatrix} \in \mathbb{R}^{(m+p)2\tau_\mathrm{f} },
\end{align*}
while the constraints for TPC are
\begin{align*}
    G &=
    \begin{bmatrix}
        \hat{H}_u \\
        -\hat{H}_u
    \end{bmatrix} \in \mathbb{R}^{2p  \tau_\mathrm{f} \times m  \tau_\mathrm{f}}, \\
    \textbf{h} &= 
    \begin{bmatrix}
        \textbf{y}_\mathrm{ub} - \hat{H}_\mathrm{p} \textbf{z}_\mathrm{p} \\
        -\textbf{y}_\mathrm{ub} + \hat{H}_\mathrm{p} \textbf{z}_\mathrm{p}
    \end{bmatrix} \in \mathbb{R}^{2p  \tau_\mathrm{f} \times m  \tau_\mathrm{f}}, \\
     \textbf{x}_\mathrm{lb} &= 
    \begin{bmatrix}
        \textbf{u}_\mathrm{lb} \\
    \end{bmatrix} \in \mathbb{R}^{m  \tau_\mathrm{f} }, \\
      \textbf{x}_\mathrm{ub} &= 
     \begin{bmatrix}
        \textbf{u}_\mathrm{ub}
    \end{bmatrix} \in \mathbb{R}^{m  \tau_\mathrm{f} }.
\end{align*}
PIQP \cite{piqp} is a second-order primal–dual interior-point method, whereas OSQP \cite{osqp} is a first-order operator-splitting solver which employs the Alternating Direction Method of Multipliers (ADMM) \cite{boyd2011distributed}, which is equivalent to applying Douglas–Rachford splitting \cite{douglas1956numerical,lions1979splitting} to the dual quadratic program.

\subsection*{Conic Program Solvers: Clarabel and SCS}
Taking the following quadratic-cost conic program form,
\begin{subequations}
\begin{align*}
    \min\limits_{\textbf{x}} \quad & \frac{1}{2} \textbf{x}^\top P \textbf{x} + \textbf{q}^\top \textbf{x} \\
    \text{s.t.} \quad & A \textbf{x} + \textbf{s} = \textbf{b}, \\
     & \textbf{s} \in \mathcal{K},
\end{align*}
\end{subequations}
the matrices for DeePC are
\begin{align*}
    A &=
    \begin{bmatrix}
        U_\mathrm{p} & 0 \\
        Y_\mathrm{p} & -I \\
        U_\mathrm{f} & 0 \\
        Y_\mathrm{f} & 0 \\
        -U_\mathrm{f} & 0 \\
        -Y_\mathrm{f} & 0 \\
    \end{bmatrix} \in \mathbb{R}^{(m+p)(\tau_\mathrm{p} + 2\tau_\mathrm{f}) \times (n_\mathrm{col} + p  \tau_\mathrm{p})}, \\
    \textbf{b} &= 
    \begin{bmatrix}
        \textbf{u}_\mathrm{p} \\
        \textbf{y}_\mathrm{p} \\
        \textbf{u}_\mathrm{ub} \\
        \textbf{y}_\mathrm{ub} \\
        -\textbf{u}_\mathrm{lb} \\
        -\textbf{y}_\mathrm{lb} \\
    \end{bmatrix} \in \mathbb{R}^{(m+p)(\tau_\mathrm{p} + 2\tau_\mathrm{f}) },
\end{align*}
while the matrices for TPC are
\begin{align*}
    A &=
    \begin{bmatrix}
        I \\
        \hat{H}_u \\
        -I \\
        -\hat{H}_u \\
    \end{bmatrix} \in \mathbb{R}^{(m+p)2\tau_\mathrm{f} \times m  \tau_\mathrm{f}}, \\
    \textbf{b} &= 
    \begin{bmatrix}
        \textbf{u}_\mathrm{ub} \\
        \textbf{y}_\mathrm{ub} - \hat{H}_\mathrm{p} \textbf{z}_\mathrm{p} \\
        -\textbf{u}_\mathrm{lb} \\
        -\textbf{y}_\mathrm{ub} + \hat{H}_\mathrm{p} \textbf{z}_\mathrm{p} \\
    \end{bmatrix} \in \mathbb{R}^{(m+p)2\tau_\mathrm{f} }.
\end{align*}
For a given row of $A \textbf{x} + \textbf{s} = \textbf{b}$, the cone $\mathcal{K}$ is the zero cone if the row is an equality, or the positive cone if the row is an inequality \cite{scs2021}.

Both Clarabel \cite{clarabel} and SCS \cite{scs2021} employ homogeneous embeddings of the KKT conditions (though the embeddings differ). 
Clarabel is a second-order primal–dual interior-point method, while SCS is a first-order Douglas–Rachford splitting method.


\begin{figure*}[htb]
    \centering
    \includegraphics[width=1.0\linewidth]{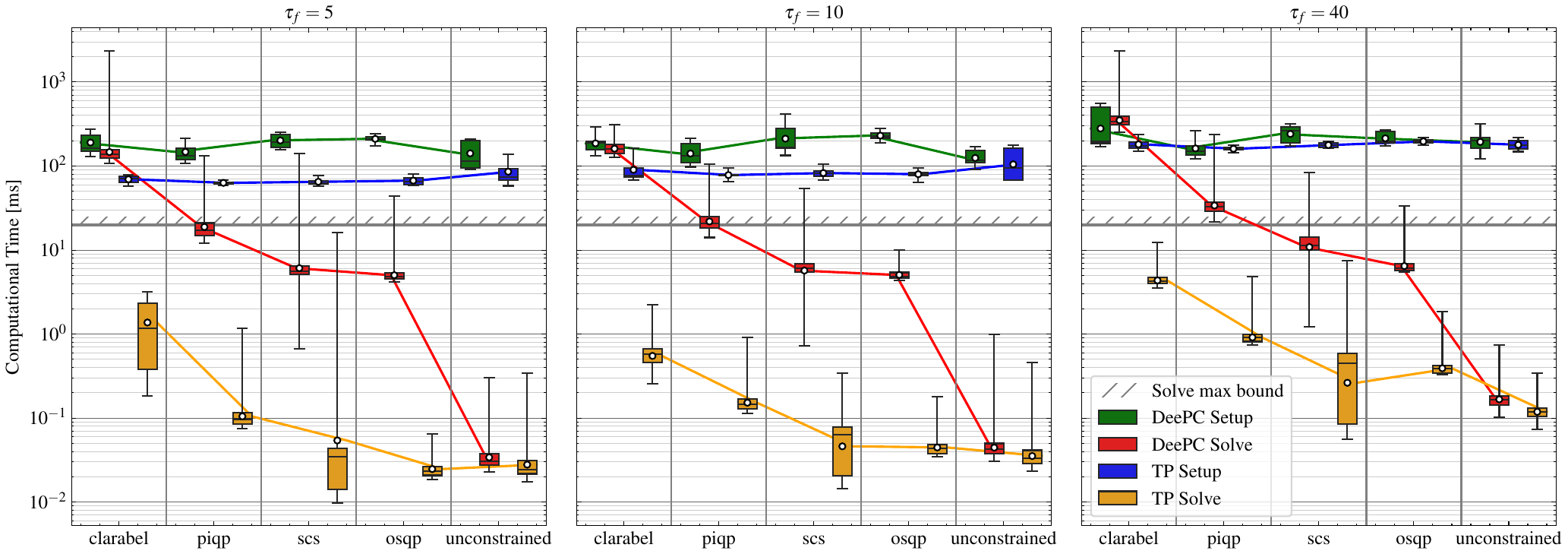}
    \caption{Comparison of different optimization formulations for problems \eqref{eq:deepc} and \eqref{eq:tpc} for different prediction lengths $\tau_\mathrm{f}$. ``unconstrained'' corresponds to the 2-norm closed-form solutions \eqref{eq:DeePCclosedForm} and \eqref{eq:unconstTPC}.}
    \label{fig:results_solvers_time}
\end{figure*}

\subsection{Closed-form Solutions for the Unconstrained Case}
If the input and output constraints of \eqref{eq:deepc} or \eqref{eq:tpc} are ignored, we get closed-form, matrix-vector product solutions for the optimal input. We include them here to compare with the computational cost of the constrained optimizations described in Appendix \ref{sec:constrainedOCPformulations}.

\subsubsection*{DeePC}
The standard quadratic \eqref{eq:deepc} does not have a closed form solution as $U_\mathrm{p} \textbf{g} = \textbf{u}_\mathrm{p}$ must be taken into account. To create an unconstrained formulation that can be solved in closed-form, we make the following alteration to \eqref{eq:deepc},
\begin{align}
    \min_{\textbf{g}, \pmb{\sigma}} \quad & \| \textbf{y} \|_{Q}^2 + \| \textbf{u} \|_{R}^2 + \lambda_{g_2} \| \textbf{g} \|_2^2 + \lambda_\sigma \| \pmb{\sigma} \|_2^2 \\
    \text{s.t.} \quad &
    \begin{bmatrix}
        Z_\mathrm{p} \\
        U_\mathrm{f} \\
        Y_\mathrm{f} \\
    \end{bmatrix} \textbf{g} =
    \begin{bmatrix}
        \textbf{z}_\mathrm{p} + \pmb{\sigma} \\
        \textbf{u} \\
        \textbf{y} \\
    \end{bmatrix}, 
    \tag{modified DeePC}\label{eq:modified-deepc}
\end{align}
where $Z_\mathrm{p}$ is the vertical concatenation of $U_\mathrm{p}$ and $Y_\mathrm{p}$, and $\textbf{z}_\mathrm{p}$ is the concatenation of $\textbf{u}_\mathrm{p}$ and $\textbf{y}_\mathrm{p}$.
Substituting, we get
\begin{align*}
    \min_{\textbf{g}} \quad & \| Y_\mathrm{f} \textbf{g} \|_{Q}^2 + \| U_\mathrm{f} \textbf{g} \|_{R}^2 + \lambda_{g_2} \| \textbf{g} \|_2^2 + \lambda_\sigma \| Z_\mathrm{p} \textbf{g} - \textbf{z}_\mathrm{p} \|_2^2,
\end{align*}
for which the optimal $\textbf{g}^\star$, given by differentiating with respect to $\textbf{g}$, is
\begin{align*}
    \textbf{g}^\star &= \lambda_\sigma  \big[Y_\mathrm{f}^\top Q Y_\mathrm{f} + U_\mathrm{f}^\top R U_\mathrm{f} + \lambda_{g_2} I + \lambda_\sigma Z_\mathrm{p}^\top Z_\mathrm{p} \big]^\dagger  Z_\mathrm{p}^\top \textbf{z}_\mathrm{p},
\end{align*}
which gives the optimal input 
\begin{align}\label{eq:DeePCclosedForm}
    \! \textbf{u}^\star \!\! = \! \lambda_\sigma U_\mathrm{f} \big[Y_\mathrm{f}^\top Q Y_\mathrm{f} \! + \! U_\mathrm{f}^\top R U_\mathrm{f} \! + \! \lambda_{g_2} I \! + \! \lambda_\sigma Z_\mathrm{p}^\top Z_\mathrm{p} \big]^\dagger Z_\mathrm{p}^\top
    \textbf{z}_\mathrm{p}.
\end{align}
Thus, the closed form solution of \eqref{eq:modified-deepc} 
is a matrix-vector product with the previous inputs and outputs. 

\subsubsection*{TPC}
The 2-norm-cost, unconstrained form of TPC is 
\begin{align*} 
    \min_{\textbf{u}} \quad & \| \hat{H}_\mathrm{p} \textbf{z}_\mathrm{p} + \hat{H}_u \textbf{u} \|_{Q}^2 + \| \textbf{u} \|_{R}^2 
\end{align*}
Differentiating with respect to \textbf{u}, we get the following closed-form solution for the optimal, unconstrained TPC trajectory:
\begin{align}
    \textbf{u}^\star &= -(\hat{H}_u^\top Q \hat{H}_u + R)^{-1}  \hat{H}_u^\top Q \hat{H}_\mathrm{p} \textbf{z}_\mathrm{p}, \label{eq:unconstTPC}
\end{align}
which is also a matrix-vector product with $\textbf{z}_\mathrm{p}$.

\subsection{Computation Comparison}
\autoref{fig:results_solvers_time} depicts the computation time in milliseconds for the selected solvers across varying prediction horizon lengths $\tau_\mathrm{f}$, separated into setup and solve phases. The solve phase explicitly exploits the solvers' warm-start capabilities and their respective \texttt{update()} interfaces.

As illustrated, OSQP and SCS significantly outperform the other solvers on the OCPs \eqref{eq:deepc} and \eqref{eq:tpc}, with OSQP exhibiting lower variance in solve time. This performance gap stems from fundamental algorithmic differences: OSQP and SCS are first-order operator-splitting methods that effectively leverage warm-starting. In contrast, PIQP and Clarabel are interior-point methods whose convergence relies on following a central path, which inherently limits the effectiveness of warm-starting. 

Remarkably, OSQP and SCS achieve solve times for the constrained problem \eqref{eq:tpc} that are comparable to the unconstrained closed-form solution that is a matrix--vector multiplication \eqref{eq:unconstTPC}. This efficiency arises because OSQP reuses the pre-factorized KKT system across solves. 
With solve times on the order of $10^{-2}~\si{\milli\second}$, the execution of \eqref{eq:tpc} falls comfortably within real-time requirements, demonstrating the viability of this data-driven approach even when enforcing convex constraints.

\end{document}